\documentclass[final,4p,authoryear]{elsarticle}
\usepackage{amssymb}
\usepackage{natbib}
\usepackage{epstopdf}
\usepackage{amscd}
\usepackage{amsmath}
\usepackage{amsfonts}
\usepackage{bm}
\usepackage{bbm}
\usepackage{amssymb}
\usepackage{color}
\usepackage{latexsym}
\usepackage{mathrsfs}
\usepackage{amsthm}
\usepackage{mathrsfs}
\usepackage{booktabs}
\usepackage{caption}
\usepackage{url}
\usepackage{rotating}
\usepackage{multirow}
\usepackage{subfig} %note: subfig must be included after the `caption` package.
\usepackage{adjustbox}
\usepackage{blindtext}
\biboptions{round}

\def\bb{\mathbf{b}}

\def\0{\mbox{\bf{0}}}

\def\bx{\mathbf{x}}

\def\bdelta{\mbox{\boldmath $\delta$}}

\def\bxi{\mbox{\boldmath $\xi$}}

\def\bXi{\mathbf{\Xi}}
\def\sE{\mathsf{E}}

\def\sV{\mathsf{V}}

\def\sS{\mathsf{S}}
\def\sI{\mathsf{I}}

\def\sM{\mathsf{M}}

\def\sN{\mathsf{N}}

\def\vec{\mbox{vec}}
    
\def\trasp{\mathsf{T}}

\newcommand{\xp}{\mathbb{E}}

\definecolor{darkross}{rgb}{0.008,0.412,0.471}
\definecolor{middleross}{rgb}{0.012,0.580,0.663}
\definecolor{lightross}{rgb}{0.016,0.749,0.855}
\definecolor{darkblue}{rgb}{0.067,0.008,0.471}
\definecolor{middleblue}{rgb}{0.094,0.012,0.663}
\definecolor{lightblue}{rgb}{0.122,0.016,0.855}
\definecolor{darkpurple}{rgb}{0.471,0.008,0.412}
\definecolor{middlepurple}{rgb}{0.663,0.012,0.580}
\definecolor{lightpurple}{rgb}{0.855,0.016,0.749}
\definecolor{darkbrown}{rgb}{0.471,0.067,0.008}
\definecolor{middlebrown}{rgb}{0.663,0.094,0.012}
\definecolor{lightbrown}{rgb}{0.855,0.122,0.016}
\definecolor{darkolive}{rgb}{0.412,0.471,0.008}
\definecolor{middleolive}{rgb}{0.580,0.663,0.012}
\definecolor{lightolive}{rgb}{0.749,0.855,0.016}
\definecolor{darkgreen}{rgb}{0.008,0.417,0.067}
\definecolor{middlegreen}{rgb}{0.012,0.663,0.094}
\definecolor{lightgreen}{rgb}{0.016,0.855,0.122}
\definecolor{darkocre}{rgb}{0.471,0.298,0.008}
\definecolor{middleocre}{rgb}{0.663,0.420,0.012}
\definecolor{lightocre}{rgb}{0.855,0.541,0.016}
\def\colorred{\color{red}}

\def\qmo{``}
\def\qmc{''}
\def\qmcsp{'' }

\newcommand\bbone{\ensuremath{\mathbbm{1}}}

\begin{document}

\begin{frontmatter}



\title{Are news important to predict large losses?}


\author[uniroma1--memotef]{M. Bernardi}
\author[uniroma2--torvergata]{L. Catania}
\author[uniroma1--memotef]{L. Petrella\corref{cor1}}
\address[uniroma1--memotef]{Department of Methods and Models for Economics, Territory and Finance,  Sapienza University of Rome}
\address[uniroma2--torvergata]{Department of Economics and Finance, University of Rome Tor Vergata}

\cortext[cor1]{Corresponding author, MEMOTEF Department, Sapienza University of Rome, Via del Castro Laurenziano, 9, 00166 Rome, e-mail: \texttt{lea.petrella@uniroma1.it}}

\begin{abstract}
In this paper we investigate the impact of news to predict extreme financial returns using high frequency data. We consider several model specifications differing for the dynamic property of the underlying stochastic process as well as for the innovation process.
%%%
Since news are essentially qualitative measures, they are firstly transformed into quantitative measures which are subsequently introduced as exogenous regressors into the conditional volatility dynamics.
%%%
Three basic sentiment indexes are constructed starting from three list of words defined by historical market news response and by a discriminant analysis.
%%%
Models are evaluated in terms of their predictive accuracy to forecast out--of--sample Value--at--Risk of the STOXX Europe 600 sectors at different confidence levels using several statistic tests and the Model Confidence Set procedure of Hansen \textit{et al.} \citeyearpar{hansen_etal.2011}.
%%%
Since the Hansen's procedure usually delivers a set of models having the same VaR predictive ability, we propose a new forecasting combination technique that dynamically weights the VaR predictions obtained by the models belonging to the optimal final set.
%%%
Our results confirms that the inclusion of exogenous information as well as the right specification of the returns' conditional distribution significantly decrease the number of actual versus expected VaR violations towards one, as this is especially true for higher confidence levels.
\end{abstract}


\end{frontmatter}

% \linenumbers

%:::::::::::::::::::::::::::::::::::::::::::::::::::::::::::::::::::::
% SECTION: INTRODUCTION
%:::::::::::::::::::::::::::::::::::::::::::::::::::::::::::::::::::::
\section{Introduction}
\label{sec:intro}
%:::::::::::::::::::::::::::::::::::::::::::::::::::::::::::::::::::::
%
\noindent The interaction between news and market volatility plays an important role in risk management and asset pricing. Indeed, there are no doubts that information about individual institutions' health conditions as well as the general macroeconomic situation move investors decisions. As theorised by the efficient market hypothesis, a direct relation may exist between news arrival and the conditional variance of observed stock returns. More precisely, looking at the semi--strong form of the \qmo Efficient Market Hypothesis\qmcsp (EMH) of Fama \citeyearpar{fama.1969}, stock prices adjust to publicly available news information very rapidly and in an unbiased fashion. The \qmo Mixture of Distribution Hypothesis\qmcsp (MDH) of Clark \citeyearpar{clark.1973} also points out a direct linkage between risk and information, arguing that there exists a serially correlated mixing variable, measuring the rate at which information arrives to the market, that drives the volatility of stock returns.
%%%
The enduring interest in return predictability using news information reflects its enormous implications. For practitioners, predictable asset returns can affect the asset allocation and hedging decisions of investors and the timing of security issuances by firms. For academics, the existence of predictable returns has implications for market efficiency and for how aggregate fluctuations in the economy are transmitted to and from financial markets.
%%%
Moreover, recent advances in asset pricing theory and a huge empirical literature have provided a strong evidence that conditional variance of stock prices is predictable, to some extent, using publicly available information. Numerous recent papers provide evidence that the predictability results from business cycle movements and changes in investors' perceptions of risk that are reflected in time--varying risk premiums. Others, however, provide evidence that predictability reflects an inefficient market populated with overreacting and irrational investors (see e.g. Mitra and Mitra, \citeyear{mitra_mitra.2011}; Kalev \textit{et al.}, \citeyear{kalev_etal.2011}; Chopra \textit{et al.}, \citeyear{chopra_etal.1992}; De Bondt and Thaler, \citeyear{debondt_thaler.1985}; Lehmann, \citeyear{lehmann.1990}). \newline
%%%
\indent A related unstressed important question, which has not been deeply investigated, is whether the inclusion of news as exogenous regressors helps to forecast extreme returns frequently observed in financial time series sampled at high frequencies. This paper adopts this point of view and investigates the impact of information to predict extreme financial returns using high frequency data in an extensive study conducted on nineteen sectors built from the STOXX Europe 600 constituents according to the \qmo FactSet\qmcsp classification.
%%%
By extreme financial returns we mean events that adversely affect the value of investor's assets or portfolios, violating the Value--at--risk (VaR) measure at 1\% confidence level.
%%%
%Jorion \citeyearpar{jorion.2007} defines the VaR as a measure of the highest expected loss, over a given time interval, under normal market conditions, at a given confidence level.
From a risk management point of view, more accurate VaR forecasts result in several benefits such as a better prediction of future losses and less capital to hold for regulatory requirements as pointed out in Basel \citeyearpar{basel.1996}. Moreover, since the VaR forecast is often linked with the volatility prediction, %it follows that also other financial applications such as derivatives and asset pricing, can benefit from the inclusion of news information.\newline
it follows that also derivative and asset pricing, can benefit from more accurate VaR predictions.\newline
%from the inclusion of news information.\newline
%%%
\indent To evaluate the relevance of the news information to predict the conditional distribution of large returns, we include exogenous covariates into the specification of several dynamic volatility models differing for the dynamic property of the underlying stochastic process as well as for the innovation assumption.
%%%
Dynamic conditional volatility models have become one of the most popular risk management tool for modelling the variability of stock returns. For our empirical analysis we refer to the class of autoregressive conditional heteroskedastic (ARCH) models introduced by Engle \citeyearpar{engle.1982}. ARCH--type models of Engle \citeyearpar{engle.1982} and Bollerslev  \citeyearpar{bollerslev.1986} offer a simple and effective method to empirically investigate the relevance of including exogenous regressors accounting for the news.\newline %\newline %as exogenous regressors in the conditional volatility process.\newline
%%%
\indent When dealing with information, a first challenge is to find an appropriate proxy for the news process. The intrinsic qualitative nature of news implies the necessity of a model which transforms this stream of information into quantitative measures.
%The intrinsic qualitative nature of news implies that a model which transforms this steams of qualitative information into quantitative measures, becomes necessary.
%
Although seminal studies use volumes and quotes as proxies for news (see e.g. Lamoureux and Lastrapes, \citeyear{lamoureux_lastrapes.1990}; Ito and Roley, \citeyear{ito_roley.1986}), many researchers argues that these kind of proxies are biased and are not appropriated, see e.g. Kalev \textit{et al.} \citeyearpar{kalev_etal.2011}. More recent studies (see Hafez \citeyear{hafez.2009}, which refers to RavenPack's technique) use alternative proxies which account also for some measures of goodness or badness which are normally referred to \qmo sentiment\qmcsp indicators. %Good sentiment indicators try to move from text to context, or even, to understand what a phrase is really saying and not simply consider a word--by--word approach.
For a comprehensive and up to date survey on news analytics techniques, see Das \citeyearpar{das.2011}.\newline
%%%
\indent Througout this paper the empirical investigation is conduced using five regressors: the lagged volumes, the number of words in news headlines referred to specific companies and three basic sentiment indicators which accounts for \qmo good\qmc, \qmo bad\qmc, and \qmo high\qmcsp number of volatility words in companies related news headlines. Moreover, since high frequency data are necessary in order to link the news to extreme financial losses, we use five minutes data to build exogenous regressors.\newline
%
%These latter indicators are company--specific and macro--related, scheduled (such as earnings report, conference call, central banks or politics announcement) and unscheduled to have a complete picture of the information flow.
%
%Another important aspect to deal with in building exogenous regressors accounting for the news arrival process is the time frame frequency.
%
%In this study exogenous regressors are built using 5--minutes returns. High frequency data are necessary in order to link the news to extreme financial losses.\newline
%High frequency data and time stamped news are necessary in order to obtain a good model.
%
%Because of the characteristics of the news process, testing the same model on a daily time frame would report poor results because of the very low market time reaction about news, in particular if these are scheduled.\newline
%%%
\indent Several alternative models obtained by combining the conditional volatility specification, the different exogenous regressors accounting for news, and the different error distribution assumption, are considered.
%%%
Forecasted VaRs delivered by competing models are then backtested using several criteria. We use the actual over expected exceedance ratio, the mean and the maximum absolute deviations of violating return of McAleer and da Veiga \citeyearpar{mcaleer_daveiga.2008}, the conditional  and unconditional coverage test of Christoffersen \citeyearpar{christoffersen.1998} and Kupiec \citeyearpar{kupiec.1995}, and the dynamic quantile (DQ) test of Engle and Manganelli \citeyearpar{engle_manganelli.2004}. As pointed out by Chen \textit{et al.} \citeyearpar{chen_etal.2012} these measures go beyond assessing violation rates and allow risk management to incorporate loss magnitudes.
%%%
Despite their recognised relevance to compare VaR violations, those tests fail to discriminate alternative models on the basis of the predictive accuracy of the VaRs. For model selection purposes, we consider the Model Confidence Set procedure (MCS) recently developed by Hansen \citeyearpar{hansen.2003}, Hansen and Lunde \citeyearpar{hansen_lunde.2005}, and Hansen \textit{et al.} \citeyearpar{hansen_etal.2011} adapted to the case where VaR forecasts' performance have to be assessed.
%%%
The Hansen's procedure consists of a sequence of statistic tests which permits to construct a set of \qmo superior\qmcsp models (SSM), where the null hypothesis of equal predictive ability (EPA) is not rejected. Since our goal is to obtain a SSM consisting of models that are statistically equivalent to predict future VaR levels, throughout the model selection procedure we propose to use a loss function explicitly tailored to penalise returns above and below the forecasted VaR in an asymmetric way. The loss function has been previously used by Gonzales \textit{et al.} \citeyearpar{gonzalez_etal.2004} for for backtesting purposes.\newline
%%%
%The  MCS procedure delivers a set of models with superior predictive ability for which the forecasted Value--at--Risk are available.
%%
\indent The presence of several models with the same predictive power in terms of the VaR forecasting, open the question of pooling the information contained in each model forecast. Since the seminal works of Reid (\citeyear{reid.1968}, \citeyear{bates_granger.1969}) on forecast combinations, several motivations have been put forward to justify such technique. Stock and Watson (\citeyear{stock_watson.1999}, \citeyear{stock_watson.2001}, \citeyear{stock_watson.2004}), for example, found that combining forecasts usually produces better results than simpler model selection techniques. Hoeting \textit{et al.} \citeyearpar{hoeting_etal.1999} deal with \qmo model uncertainty\qmcsp in a Bayesian framework by averaging different model specifications. Within a decision--theoretical approach, combining forecasts can be seen as a procedure aiming at diversifying and then reducing the risk of selecting a single model to produce forecasts. The recent surveys of Timmermann \citeyearpar{timmermann.2006} and Aiolfi \textit{et al.} \citeyearpar{aiolfi_etal.2011} provide further motivations for combining forecasts.\newline
%%%
\indent In this paper, we propose a dynamic procedure that combines VaR forecasts delivered by the models belonging to the superior set. As argued by Giacomini and Komunjer \citeyearpar{giacomini_komunjer.2005}, the combination for VaR models is beneficial since the VaR is a small coverage quantile model which is sensitive to the few observations below the quantile estimate. Moreover, combining different VaR models can also robustify individual VaR forecasts, since estimated model parameter are particularly unstable over time. We propose to combine VaR predictions by a dynamic weighting scheme using a new exponential smoothing moving average process. The smoothing process we propose for the weights considers an asymmetric innovation distribution accounting also for the parameter uncertainty problem.\newline
%
%Combining different VaR models can robustify individual VaR forecasts, since estimated model parameter are particularly unstable over time. In this respect, VaR combination, in general, and the dynamic procedure here considered, also accommodate the parameter uncertainty problem.\newline
%%
%In this respect, VaR combination, in general, and the dynamic procedure here considered, also accommodate the parameter uncertainty problem.\newline
%%
%which is particularly useful when the MCS procedure is not able to further discriminate models. We dynamically weight the VaR predictions delivered by the models belonging to optimal set using an exponential smoothing moving average process. The dynamic VaR combination can be seen as a method to optimally choose the model weights as the time goes by. As argued by Giacomini and Komunjer \citeyearpar{giacomini_komunjer.2005}, the combination for VaR models is beneficial since the VaR is a small coverage quantile model which is sensitive to the few observations below the quantile estimate. For this reason the smoothing process we propose for the weights consider an asymmetric innovation distribution. Combining different VaR models can robustify individual VaR forecasts, since estimated model parameter are particularly unstable over time. In this respect, VaR combination, in general, and the dynamic procedure here considered, also accommodate the parameter uncertainty problem.\newline
%
\indent Our empirical results confirm that the information news significantly improves the prediction of extreme values in an extensive out--of--sample analysis. Moreover, we show that the combination of an \qmo ad hoc\qmcsp model selection procedure with an optimal dynamic VaR combination technique provides a better VaR prediction. Finally, as additional consideration, we find that the introduction of exogenous regressors explains the seasonal volatility patterns up to a weekly period which characterise high frequency financial data, see Gencay \textit{et al.} \citeyearpar{gencay_etal.2001}.\newline
%
%::::::::::::::::::::::::::::::::::::::::::::::::::::::::::::::::::::::
% Organization of the paper
%::::::::::::::::::::::::::::::::::::::::::::::::::::::::::::::::::::::
%
\indent The remaining of the paper is organised as follows. In Section \ref{sec:literature_review} we systematically review the empirical literature on the relevance of news to predict financial returns. Section \ref{sec:model} details the model specifications, while Section \ref{sec:model_sel} describes the Hansen \textit{et al.} \citeyearpar{hansen_etal.2011} model selection procedure. Section \ref{sec:combining_vars} describes the dynamic VaR combination technique developed to pool VaR forecasts. Section \ref{sec:empirical_analysis} describes the data, provides insights about the construction of exogenous regressors, and discusses the main empirical results. Section \ref{sec:conclusion} concludes.
%
%::::::::::::::::::::::::::::::::::::::::::::::::::::::::::::::::::::::
% Section: Literature review
%::::::::::::::::::::::::::::::::::::::::::::::::::::::::::::::::::::::
\section{Literature review}
\label{sec:literature_review}
%::::::::::::::::::::::::::::::::::::::::::::::::::::::::::::::::::::::
%
\noindent Since the seminal work of Clark \citeyearpar{clark.1973}, which is normally referred to as the Mixture of Distribution Hypothesis (MDH), a relevant number of research works appeared in the financial econometric literature focusing on the relationship between stock's volatility and news, one of  the major problem being the choice of an adequate proxy for the news arrival process. The first natural answer to this relevant question was to use volumes and quotes (see e.g. Ito and Roley \citeyear{ito_roley.1986} and Lamoureux and Lastrapes \citeyear{lamoureux_lastrapes.1990}). The recent work of Kalev \textit{et al.} \citeyearpar{kalev_etal.2004} argues that there are three main factors to be taken into account using trading volumes as proxy for the informations: the endogenous nature of volumes, the presence of different kind of traders having different news interpretations (for example, noise trader versus liquidity trader), and the fact that volumes could not account for private information.
%%%
Another stream of literature in early 90's focused instead on news classification. Jones \textit{et al.} \citeyearpar{jones_etal.1998}, for example, used scheduled versus unscheduled news in a EGARCH model finding significative results modelling the first two moments of the USD/AUD exchange rate. Many other studies instead model the impact of news in the conditional volatility equation using specific dummy variables: Blasco \textit{et al.} \citeyearpar{blasco_etal.2002} use four dummies (respectively for a combination of good--bad news and high--low relevance news) in a GJR--GARCH specification showing that bad news could be used to capture the asymmetric volatility response of negative shocks. The related work of Kim \citeyearpar{kim.1998} uses similar proxies in a EGARCH model for the USD/AUD exchange rate finding significative results even for the conditional mean specification. In a multivariate framework, Fornari \textit{et al.} \citeyearpar{fornari_etal.2002} used newspaper headlines as exogenous variables in a Markov switching vector autoregressive model with GARCH innovations. As a consequence of the technology innovation and computer power improvement, the last decade has seen a rapid increase of the number of works focusing on the use of publicly available information. DeGennaro and Shrieves \citeyearpar{degennaro_shrieves.1997} and Ederington and Lee \citeyearpar{ederington_lee.1993} use the number of macroeconomic news to explain foreign exchange conditional volatility, a technique that in recent years is referred to the naive classification (see also Das \citeyearpar{das_chen.2007}, for an extensive survey on news analytics technique).
%%%
The use of more sophisticated news analytic technique, gives rise to the big challenge of moving from text to context. In other words, this means that it is really difficult to quantify the message behind new headline stamps.
%
%The biggest challenge in using more sophisticated news analytic technique, is to move from text to context. In other words, it is really difficult to quantify the message behind words.
%
The common approach is to create an index accounting for some news measures. This index could be a simple news--related words counter or a more sophisticated sentiment index (see Hafez \citeyear{hafez.2009}, which refers to RavenPack's technique). Kalev and Duong \citeyearpar{kalev_etal.2011} and Kalev \textit{et al.} \citeyearpar{kalev_etal.2004} use an EGARCH model where the number of news enters the conditional volatility equations as exogenous regressor, to explain intradaily volatility patterns of the S\&P/ASX 200 Index. They find that the rate of information arrivals has a positive impact on volatility. Mitchell and Harold \citeyearpar{mitchell_harold.2009} provide similar results using a weekly timeframe, while Berry and Howe \citeyearpar{berry_etal.1994} do not find any significant relation between number of news releases by Reuter's News Service and the S\&P500 index volatility.\newline
\indent In this paper, we approach the problem of quantifying the impact of news information from the different point of view of measuring extreme financial losses. To this end, we employ the word counter approach, i.e. the sentiment indexes are function of the number of specific words class, using a new information supplier (the FactSet workstation), and the new classification of \qmo high\qmcsp words associated with the the well know \qmo good, bad\qmcsp scheme.
%We decide to consider different dynamics for the conditional variance and to employ the Model Confidence Set of Hansen \textit{et al.} \citeyearpar{hansen_etal.2011} and Hansen \citeyearpar{hansen.2003} to select a set of models that are equals in terms of their predictive ability. Our results is that information could improve a Value--at--Risk backtest using high frequency data.
%
%::::::::::::::::::::::::::::::::::::::::::::::::::::::::::::::::::::::
% SECTION: THE MODEL
%::::::::::::::::::::::::::::::::::::::::::::::::::::::::::::::::::::::
\section{Model specifications}
\label{sec:model}
%::::::::::::::::::::::::::::::::::::::::::::::::::::::::::::::::::::::
%
\noindent As detailed in the Introduction, in this paper we investigate and compare VaR forecasts obtained using a list of popular autoregressive conditional heteroskedastic (ARCH) models introduced by Engle \citeyearpar{engle.1982} and subsequently generalised to the GARCH family by Bollerslev \citeyearpar{bollerslev.1986}. In order to account for the well known stylised facts about financial returns we consider several specifications differing for the conditional volatility dynamics as well as the distributions of the error term. Furthermore, a first--order autoregressive term is included into the conditional mean specification of all the competing models to capture the negative first order autocorrelation frequently displayed by high frequency data, as documented by Gencay \textit{et al.} \citeyearpar{gencay_etal.2001} among others.\newline
%%%
\indent Formally, let $r_t$ the logarithmic return at time $t$, for $t=1,2,\dots,T$, we consider the following general model specification
\begin{eqnarray}
r_t=\mu+\phi r_{t-1}+\varepsilon_t,\qquad\varepsilon_t=\sigma_t\zeta_t,\qquad\zeta_t\sim\mathcal{D}\left(0,1\right),\nonumber
\end{eqnarray}
where $\zeta_t$ is a sequence of independently and identically distributed random variables with general distribution $\mathcal{D}\left(0,1\right)$ having mean zero and variance 1, $\sigma_t$ is the conditional standard deviation of $r_t$ and $\phi$ is the autoregressive parameter assumed to be $\vert\phi\vert<1$ to preserve stationarity.
%%%
Regarding the specification of the error distribution $\mathcal{D}\left(0,1\right)$, we consider the usual thin--tailed Gaussian $\mathcal{N}\left(0,1\right)$ assumption and the two alternatives, i.e. the Student--t $\mathcal{T}_{\nu}\left(0,\frac{\nu-2}{\nu}\right)$, with $\nu$ degrees of freedom and the Generalised Error distribution (GED), being able to reproduce the high level of kurtosis frequently observed empirically. The standardised GED, i.e. $\mathcal{GE}\left(0,1,\nu\right)$ has conditional density given by
%%%
\begin{equation}
f\left(0,1,\nu\right)=\frac{\nu\exp\left\{-\frac{1}{2}\left(\sqrt{2^{2/\nu}\frac{\Gamma\left(\nu^{-1}\right)}{\Gamma\left(3\nu^{-1}\right)}}\vert\epsilon_t\vert\right)^\nu\right\}}{\sqrt{2^{2/\nu}\frac{\Gamma\left(\nu^{-1}\right)}{\Gamma\left(3\nu^{-1}\right)}}
2^{1+\nu^{-1}\Gamma\left(\nu^{-1}\right)}}\bbone_{\left(-\infty,\infty\right)}\left(\epsilon_t\right),
\end{equation}
%%%
where the shape parameter $\nu\in\left(0,\infty\right)$ regulates the tail behaviour: $\nu=2$ corresponds to the Gaussian density while $\nu=1$ correspond to the Laplace distribution of Kotz \textit{et al.} \citeyearpar{kotz_etal.2001}.\newline
%%%
%and the Normal Inverse Gaussian distribution $\mathcal{NIG}\left(0,1,\nu\right)$ introduced by Barndorff--Nielsen \citeyearpar{barndorff_nielsen.1977} as a subclass of the generalised hyperbolic distribution.\newline
%
\indent We complete the model specifications by introducing a dynamics for the conditional volatility term $\sigma_t^2$ belonging to the family of GARCH models. ARCH--type models are flexible and powerful tools for conditional volatility modelling; they are able to consider the volatility clustering phenomena, the excess of kurtosis and the asymmetry of the data. Moreover, they permit to introduce exogenous information in the specification of the conditional volatility dynamics in an efficient and effective way. The simplest conditional volatility dynamics we consider as a benchmark is the GARCH(1,1) specification introduced by Bollerslev \citeyearpar{bollerslev.1986}
\begin{equation}
\sigma_t^2=\omega+\bdelta^\trasp\bx_t+\alpha\varepsilon_{t-1}^2+\beta\sigma_{t-1}^2,
\label{eq:garch_11_spec}
\end{equation}
where $\omega>0$ and $0\leq\alpha,\beta<1$ with $\alpha+\beta<1$ to preserve weak ergodic stationarity of the conditional variance. To account for asymmetric response (leverage effect, see Black \citeyear{black.1976}) we also consider two additional specifications, i.e. the EGARCH(1,1) of Nelson \citeyearpar{nelson.1991}
\begin{equation}
\log\left(\sigma_t^2\right)=\omega+\bdelta^\trasp\bx_t+g\left(\zeta_{t-1}\right)+\beta\log\left(\sigma_{t-1}^2\right),
\label{eq:egarch_11_spec}
\end{equation}
where $g\left(\zeta_{t}\right)=\alpha\zeta_t+\gamma\left(\vert\zeta_t\vert-\mathbb{E}\vert\zeta_t\vert\right)$, and the GJR-GARCH(1,1) of Glosten \textit{et al.} \citeyearpar{glosten_etal.1993}
\begin{eqnarray}
\sigma_t^2=\omega+\bdelta^\trasp\bx_t+\left(\alpha\varepsilon_{t-1}^2+\gamma\bbone_{\left(-\infty,0\right]}\left(\varepsilon_{t-1}\right)\varepsilon_{t-1}^2\right)+\beta\sigma_{t-1}^2,
\end{eqnarray}
where $\omega>0$, $0\leq\alpha,\beta,\gamma<1$ with $\alpha+\beta+\gamma\mathbb{P}\left(\zeta_t<0\right)<1$, $\mathbb{P}\left(\zeta_t<0\right)$ represents the probability to observe a negative event under the chosen distribution for $\zeta_t$, and $\bbone_{\left(a,b\right)}\left(x\right)$ denotes the indicator function having non--zero values for $x$ in the interval $\left(a,b\right)$.
%%%
In all the considered volatility specifications, $\bx_t$ denotes a vector of exogenous variables at time $t$, which measures the impact of news on the conditional volatility process, and $\bdelta$ denotes the associated regression parameter. Details on the procedure used to build exogenous quantitative regressors from recorded financial and macroeconomics news are provided in Section \ref{sec:empirical_analysis}. To estimate model parameters we consider the maximum likelihood approach, see e.g. Francq and Zakoian \citeyearpar{francq_zakoian.2010}.
%
%%%%%%%%%%%%%%%%%%%%%%%%%%%%%%%%%%%%%%%%%%%%%%%
%%% DA METTERE ALTROVE
%%%%%%%%%%%%%%%%%%%%%%%%%%%%%%%%%%%%%%%%%%%%%%%
%
%The tables summarising the parameter estimates for all the competing model specifications are available as supplementary material. The parameters associated to the conditional mean dynamic appear to be not statistically significant for some sectors; this means that for those series the well know stylised fact of negative serial autocorrelation frequently displayed by high frequency data is not present. Moreover, as described by Gencay \textit{et al.} \citeyearpar{gencay_etal.2001}, this phenomena can be associated with the procedure employed to construct the high frequency series. Also the parameter associated with the exogenous regressor sometimes appears to be not statistically significant, but this does not represent a problem for our analysis since it is mainly focused on extreme loss prediction and not on describing the whole returns behaviour. %However, we do not really care about that since our analysis is mainly focused on extreme loss prediction and not on describing the whole returns behaviour.
%
%::::::::::::::::::::::::::::::::::::::::::::::::::::::::::::::::::::::
% SECTION: MODEL SELECTION PROCEDURE
%::::::::::::::::::::::::::::::::::::::::::::::::::::::::::::::::::::::
\section{Model selection procedure}
\label{sec:model_sel}
%::::::::::::::::::::::::::::::::::::::::::::::::::::::::::::::::::::::
%
\noindent The availability of several alternative model specifications being able to adequately describe the unobserved data generating process (DGP) opens the question of selecting the \qmo best fitting model\qmcsp according to a given optimality criterion.
%
%opportunity to introduce a procedure to select the \qmo best fitting model\qmcsp according to a given optimality criterion.
%
The definition of the optimality criterion requires the prior specification of a target with respect to which the ability of each model to replicate that characteristic is evaluated. Since in this paper we are interested in comparing several model specifications on the basis of their predictive ability of large losses, a natural candidate would be the A/E violation rate or the usual conditional and unconditional coverage tests combined with a selection procedure. Despite their recognised relevance to compare VaR violations, those tests fail to discriminate alternative model specifications on the basis of the predictive accuracy of the VaRs.
%%%
%However, when dealing with extreme VaR confidence levels, such as $0.1\%$ or $1\%$, a relevant problem for the usual backtesting procedures is represented by the low number of expected violations, even with high frequency data.
%%%
To overcome this problem and to present results less sensitive to the low number of theoretical violations, we apply the Model Confidence Set (MCS) procedure recently developed by Hansen \textit{et al.} \citeyearpar{hansen_etal.2011}.
%%%
The Hansen's procedure consists of a sequence of statistic tests which permits to build the \qmo Superior Set of Models\qmcsp (SSM), where the null hypothesis of equal predictive ability (EPA) is not rejected at a certain confidence level. The EPA statistic tests requires the specification of a loss function summarising the final target of the comparison procedure. According to the target of predicting extreme losses, we propose to use as loss function the asymmetric VaR function of Gonzales \textit{et al.} \citeyearpar{gonzalez_etal.2004} which penalises for each model $j=1,2,\dots,m$ more heavily observations below the $\tau$--th quantile level, i.e. $r_{t+1}<\mathrm{VaR}^{\tau}_{j,t+1\vert t}$ than observation above it, where $\mathrm{VaR}^{\tau}_{j,t+1\vert t}$ denotes the $\tau$--level predicted VaR. In our contest, the loss function for model $j$ becomes
%%%
\begin{eqnarray}
\ell\left(r_{t+1},{\rm VaR}_{j,t+1\vert t}^{\tau}\right)=n^{-1}\rho_\tau\left(r_{t+1}-\mathrm{VaR}_{j,t+1\vert t}^{\tau}\right),
%%%
\label{eq:quantile_loss_function}
%%%
\end{eqnarray}
where $\rho_{\tau}\left(z\right)=z\left(\tau-\bbone_{\left(-\infty,0\right)}\left(z\right)\right)$, is the $\tau$--th quantile loss function and $T$ is the length of the out of sample rolling forecast.\newline\newline
%%%
\noindent We now briefly describe how the MCS procedure is implemented. The procedure starts from an initial set of models $\mathrm{M}_{0}$ of dimension $m$ encompassing all the model specifications described in Section \ref{sec:model}, and results in a hopefully smaller set of $\mathrm{\hat{M}}_{1-\alpha}^{*}$ of dimension $m^*\leq m$ for a given confidence level $1-\alpha$. The best scenario, of course, would be when the final set consists of a single mode, i.e. $m^*=1$.\newline
%%%
\indent Formally, let $d_{ij,t+1}$ denotes the loss difference between models $i$ and $j$, at time $t+1$
\begin{eqnarray}
d_{ij,t+1}=\ell\left(r_{t+1},{\rm VaR}_{i,t+1\vert t}^{\tau}\right)-\ell\left(r_{t+1},{\rm VaR}_{j,t+1\vert t}^{\tau}\right),
\end{eqnarray}
for $i,j=1,\dots,m,$ and $t=0,\dots,n-1$, under stationarity assumptions for the $d_{ij}$ series, the null hypothesis, for a given set of model is
\begin{eqnarray}
\mathrm{H}_{0}:\xp \left(d_{ij}\right)=0,\qquad\forall i,j=1,2,\dots,m
\end{eqnarray}
tested versus the alternative hypothesis
\begin{eqnarray}
\mathrm{H}_{1}:\xp\left(d_{ij}\right)\neq 0,\qquad {\rm for\,some}\quad i,j=1,\dots,m.
\end{eqnarray}
The appropriate test statistic is
\begin{equation}
\mathrm{T}_{\rm R}=\max_{i,j \in\mathrm{M}}\frac{\mid\bar d_{ij}\mid}{\sqrt{\hat{\rm var}\left(\bar d_{ij}\right)}},
\label{eq:hansen_test_statistics}
\end{equation}
defined in Hansen \textit{et al.} \citeyearpar{hansen_etal.2011}, where, $\bar{d}_{ij}=n^{-1}\sum_{t=1}^n d_{ij,t}$ measures the relative sample loss between the $i$--th and $j$--th models, while $\hat{\rm var}\left(\bar d_{ij}\right)$ is a bootstrapped estimate of ${\rm var}\left(\bar d_{ij}\right)$. According to Hansen \citeyearpar{hansen_etal.2011}, the bootstrapped variances $\hat{\rm var}\left(\bar d_{i,\cdot}\right)$, are calculated performing a block--bootstrap procedure detailed in \ref{sec:appendix_A} of this paper. The block length $p$ is chosen as the maximum number of significants autoregressive parameters resulting after fitting a stationary AR$(p)$ process on all the $d_{ij}$ series.\newline
%%%
\indent The MCS procedure sequentially eliminates the worst model until the null hypothesis of EPA is accepted for each model in the set, the elimination rule coherent with \eqref{eq:hansen_test_statistics} is
\begin{equation}
e_{\rm R}=\arg\max_{i}\left\{\sup_{j}\frac{\bar{d}_{ij}}{\sqrt{\hat{\rm var}\left(\bar d_{ij}\right)}}\right\}.
\label{eq:hansen_test_elimination_rule}
\end{equation}
Since the SSM $\hat{\sM}_{1-\alpha}^*$ delivered by the Hanen's procedure usually contains a large number of models characterised by the same VaR predictive ability, in the next section, we describe how to implement a procedure that combines the obtained VaRs forecast.
%
%
%::::::::::::::::::::::::::::::::::::::::::::::::::::::::::::::::::::::
% SECTION: COMBINING VAR FORECASTS
%::::::::::::::::::::::::::::::::::::::::::::::::::::::::::::::::::::::
\section{Combining VaRs}
\label{sec:combining_vars}
%::::::::::::::::::::::::::::::::::::::::::::::::::::::::::::::::::::::
%
\noindent In general the SSM procedure detailed in the previous section delivers, for a given confidence level $\alpha$, a set of models with superior predictive ability i.e. $\hat{\sM}_{1-\alpha}^*$, for which all the forecasted VaRs, ${\rm VaR}_{j,t+1\vert t}^{\tau}$, $\forall t=1,2,\dots,T$ and $\forall j=1,2,\dots,m^*$are available. The presence of several models with the same VaR predictive ability, open the question of pooling the information contained in each model forecast.
%%%
The theory of combining forecasts dates back to the work of Bates and Granger \citeyearpar{bates_granger.1969} and Reid (\citeyear{reid.1968}, \citeyear{reid.1969}) that inspired subsequent theoretical and empirical investigations on this topic. From the empirical point of view, Stock and Watson (\citeyear{stock_watson.1999}, \citeyear{stock_watson.2001}, \citeyear{stock_watson.2004}) find that combining forecasts usually produces better results than a simpler model selection procedure that considers only the \qmo best\qmcsp model to produce forecast. Several motivations had been put forward to justify such empirical evidence. %The main justification relies upon the fact that different model specifications provide only \qmo local approximations\qmcsp of the true unknown DGP and usually none of them dominates all the others uniformly over the period of time where the competing models are evaluated.
The main justification relies upon the consideration that each alternative model specification provides only a \qmo local approximation\qmcsp of the true unknown DGP without over performing  all the outstanding ones, uniformly over the period of time where the competing models are evaluated.
From a Bayesian perspective, the pioneer work of Hoeting \textit{et al.} \citeyearpar{hoeting_etal.1999} support the idea of dealing with \qmo model uncertainty\qmcsp by averaging different model specifications. Furthermore, within a decision--theoretical approach, combining forecasts can be seen as a procedure that aims to diversify and then reduce the \qmo model selection\qmcsp risk. The works of Timmermann \citeyearpar{timmermann.2006} and Aiolfi \textit{et al.} \citeyearpar{aiolfi_etal.2011} provide excellent surveys of recent developments on forecast combination.\newline
%%%
\indent In this work we instead consider the related topic of combining VaR forecasts at the same confidence level $\tau$. As argued by Giacomini and Komunjer \citeyearpar{giacomini_komunjer.2005}, the combination for VaR models is beneficial since the VaR is a small coverage quantile which is sensitive to the few observations below the quantile estimate. Within the general framework of conditional quantile estimate, Christoffersen \textit{et al.} \citeyearpar{christoffersen_etal.1999} provide some motivations for combining VaRs in order to obtain a single statistically superior measure but they do not give any method to select the weights associated to each model. Giacomini and Komunjer \citeyearpar{giacomini_komunjer.2005} instead construct an encompassing test to compare conditional quantile forecasts obtained from non--nested models in an out--of--sample framework. However, their work relies on a simple \qmo tick\qmcsp loss function which does not penalise more heavily observations far from the estimated quantile.
%%%
Here instead we first determine the SSM using the MCS procedure of Hansen \textit{et al.} \citeyearpar{hansen_etal.2011} based on an \qmo ad hoc\qmcsp loss function explicitly tailored to discriminate quantile--based risk measures, and then we dynamically weight the VaR predictions to produce better VaR forecasts at each point in time. The proposed dynamic VaR combination method can be seen as a method to optimally choose the model weights which also accommodates parameter uncertainty problem. Moreover, since estimated model parameters are particularly unstable over time, combining different VaR models can robustify individual VaR forecasts.\newline
\indent In particular, we propose to pool the information coming from the individual ${\rm VaR}_{j,t+1\vert t}^{\tau}$, $j=1,2,\dots,m^*$ estimates considering the following convex linear combination
\begin{eqnarray}
{\rm VaR}_{t+1\vert t}^{\tau,{\rm dyn}}=\sum_{j=1}^{m^*}\varpi_{j,t+1\vert t}{\rm VaR}_{j,t+1\vert t}^\tau,
%%%
\label{eq:var_combination}
%%%
\end{eqnarray}
where $\varpi_{j,t+1\vert t}$, for $j=1,2,\dots,m^*$ is a set of model specific weights. In principle the weights $\varpi_{j,t+1\vert t}$, for $j=1,2,\dots,m^*$ can be chosen by some optimality criterion based on the past history of the VaR violations. %We propose to weight individual VaR forecasts according to the following exponential smoothing moving average process
We propose to use the following exponential smoothing moving average process
\begin{equation}
\varpi_{t+1,j}=\kappa_j\varpi_{t,j}+\left(1-\kappa_j\right)\tilde{\pi}\left(r_t,{\rm VaR}_{j,t\vert t-1}^\tau,\hat{\sigma}_{j,t}\right),
\label{eq:vars_weighing_equation}
\end{equation}
with $\kappa_j\in\left(0,1\right)$, and $\tilde{\pi}\left(r_t,{\rm VaR}_{j,t\vert t-1}^\tau,\hat{\sigma}_{j,t}\right)$ is the exponential of the $\tau$--quantile loss kernel defined in equation \eqref{eq:quantile_loss_function} normalised over all the possible models belonging to the final set ${\rm M}_{1-\alpha}^*$, i.e.
\begin{equation}
\tilde{\pi}\left(r_t,{\rm VaR}_{j,t\vert t-1}^\tau,\hat{\sigma}_{j,t}\right)=\frac{\exp\left\{\ell\left(r_t,{\rm VaR}_{j,t\vert t-1}^\tau\right)/\hat{\sigma}_{j,t}\right\}}{\sum_j^{m^*}\exp\left\{\ell\left(r_t,{\rm VaR}_{j,t\vert t-1}^\tau\right)/\hat{\sigma}_{j,t}\right\}},
\label{eq:var_combination_weights}
\end{equation}
where $\hat{\sigma}_{j,t}$, is the predicted conditional variance at time $t$ of model $j=1,2,\dots,m^*$.
Equation \eqref{eq:var_combination_weights} coincides with the relative loss of model $j=1,2,\dots,m^*$ with respect to the weighted average loss of the VaR combinations and it guarantees an asymmetric penalisation of returns above and below the forecasted VaRs accounting also for the magnitude of VaR violations. Moreover, it can be shown that equation \eqref{eq:var_combination_weights} satisfies some optimality criteria because it is robust against model misspecifications, i.e. it minimizes the Kullback--Leibler divergence with respect to the probability density function having generated the observed data, see Sriram \textit{et al.} \citeyearpar{sriram_etal.2013}.\newline
%%%%
\indent The initial weighs at time $t=1$ are set to $\varpi_{1,j}=\frac{1}{m^*}$, for all $j=1,2,\dots,m^*$ which guarantees that $\sum_{j=1}^{m^*}\varpi_{t,j}=1$ for all $t=2,3,\dots,T$. In this way, the convex linear combination \eqref{eq:vars_weighing_equation} enforces the resulting VaR \eqref{eq:var_combination} to belong to the interval defined by the minimum and the maximum of estimated VaRs, i.e. ${\rm VaR}_{t+1\vert t}^{\tau,{\rm dyn}}\in\left(\min_j\left\{{\rm VaR}_{j,t+1\vert t}^\tau\right\},\max_j\left\{{\rm VaR}_{j,t+1\vert t}^\tau\right\}\right)$, in the optimal set $\hat{\sM}_{1-\alpha}^*$.
Finally, we estimate the autoregressive parameters $\kappa_j$ for $j=1,2,\dots,m^*$ by minimising the average asymmetric VaR loss function of Gonzales \textit{et al.} \citeyearpar{gonzalez_etal.2004} over the forecast horizon.
%%%
%The $\kappa_j$ parameters for $j=1,2,\dots,m^*$ can be chosen by some optimality criterion based on the past history of the VaR violations. We propose to chose them by minimising the average asymmetric VaR loss function of Gonzales \textit{et al.} \citeyearpar{gonzalez_etal.2004} over the forecast horizon $T$. So, let $\mathbb{\kappa}=\left(\kappa_1,\dots,\kappa_{m^*}\right)$ the set of the autoregressive parameters, after having calculated the VaR forecast series $\left(\mathrm{VaR}_1^\tau,\dots,\mathrm{VaR}_T^\tau\right)$, using Equations \eqref{eq:var_combination} and \eqref{eq:vars_weighing_equation}, it is possible to calculate the average asymmetric loss as:
%
%\begin{equation}
%\mathrm{Q}\left(\kappa,\tau,\br_1^T\right)=T^{-1}\sum_{t=1}^T \left(r_t-\mathbb{I}_{\{r_t<{\rm VaR}_{t}^\tau\}}\right)\left(r_t-{\rm VaR}_{t}^\tau\right).
%\label{eq:minimizer}
%\end{equation}
%%
%Finally the parameters $\mathbb{\kappa}=\left(\kappa_1,\dots,\kappa_{m^*}\right)$ are estimated by minimizing $\mathrm{Q}\left(\kappa,\tau,\br_1^T\right)$:
%\begin{equation}
%\hat\bkappa=\operatorname*{arg\,min}_{\bkappa}\mathrm{Q}\left(\bkappa,\tau,\br_1^T\right),
%\end{equation}
%%
%via a numerical optimiser. The resulting parameters $\mathbb{\hat\kappa}=\left(\hat\kappa_1,\dots,\hat\kappa_{m^*}\right)$ are optimal under the chosen loss function.
%
%
%::::::::::::::::::::::::::::::::::::::::::::::::::::::::::::::::::::::
% SECTION: EMPIRICAL ANALYSIS
%::::::::::::::::::::::::::::::::::::::::::::::::::::::::::::::::::::::
\section{Empirical Analysis}
\label{sec:empirical_analysis}
%::::::::::::::::::::::::::::::::::::::::::::::::::::::::::::::::::::::
%
\noindent In this section we describe how we empirically investigate the impact of the news regressors to predict large losses. In particular, we describe the procedure used to build the exogenous regressors accounting for the news arrival process. Then, we present the results concerning the application of the Hanen's MCS procedure highlighting the ability of the procedure to discriminate the different model specifications. We also discuss the results obtained using the VaR combination technique proposed in the previous section.
%
%We conclude the section discussing the results we get from the application of the VaR combination technique detailed in Section \ref{sec:combining_vars}.
%
%::::::::::::::::::::::::::::::::::::::::::::::::::::::::::::::::::::::
% SECTION: THE DATA
%::::::::::::::::::::::::::::::::::::::::::::::::::::::::::::::::::::::
\subsection{The Data}
\label{sec:data}
%::::::::::::::::::::::::::::::::::::::::::::::::::::::::::::::::::::::
%
\noindent Our dataset is made by all the STOXX Europe 600 constituents 1--minute prices and volumes from 17/08/2012 09:00:00 to 01/02/2013 09:40:00 comprising of 87120 observations per firm, recorded from the Bloomberg workstation. Starting from the 1--minute series, we construct two series at 2 and 5--minutes frequency for prices and volumes, where volumes' aggregation is made by sum. Then, we cross--sectionally aggregate individual return series into nineteen sectors according with the \qmo FactSet\qmcsp classification. Aggregation is done by averaging log--returns and by summing volumes, excluding 30 series which present missing values. The 2--minutes original dataset are used to preliminary investigate on market news response, as discussed in the following Section \ref{sec:regressors}, while the 5--minutes sectorial log--returns are used to build the model and to investigate the ability of the different models to predict large losses.\newline
%%%
\indent  The descriptive statistics for the log--returns are provided in Table \ref{tab:Sectors_return_summary_stat} for all the considered sectors. In line with stylised facts of financial time series, the returns are skewed and leptokurtic, indicating that their unconditional distribution is not Normal as documented by the Jarque--Bera (JB) statistics.\newline
%Concerning the construction of the exogenous regressor starting from qualitative information flow we consider
%%%
\indent The news information we consider come from the \qmo FactSet\qmcsp news archive which collects news from about 35 different agencies. From this database we record 81425 company related and time stamped headlines excluding duplicated ones. The headlines we consider contain various type of news: these are macroeconomics and institution specific, scheduled (such as earnings report, conference call, central banks or politics announcement) and unscheduled. The collected headlines are then cleaned from not relevant news, such as, for example, ownership update and conference calendar announces. At the end of this procedure we remain with 51266 headlines. The exogenous regressor building procedure is detailed in the next section.
%
%In the next section we provide more details on informations data and we describe the exogenous data building process.
%
 %by average the log--returns and by sum the volumes into 19 se, according with FactSet classification; 30 of the 500 series have gaps, so we decide to cut off this series. The 2--minutes original dataset was used to investigate on market news response, while the 5--minutes sectorial log-returns was used in the model. The news regressors were constructed starting from FactSet news archive; from this, we record 81425 company related and time stamped headlines. FactSet collects news from about 35 agencies; we did not consider duplicated news. Headlines contain various type of news, this are even macro and micro, schedules and unschedules; we clean our dataset from not significant news, for example ownership update and conference calendar announces, after this we remain with 51266 headlines.
%
%Good news and bad news:
%
%::::::::::::::::::::::::::::::::::::::::::::::::::::::::::::::::::::::
% SECTION: EXOGENOUS REGRESSORS
%::::::::::::::::::::::::::::::::::::::::::::::::::::::::::::::::::::::
\subsection{Exogenous Regressors and model specifications}
\label{sec:regressors}
%::::::::::::::::::::::::::::::::::::::::::::::::::::::::::::::::::::::
%
\noindent %To build the regressors and to implement the out--of--samples Value--at--Risk backtesting procedures, we split our 2--minutes dataset in two equal parts. The first part is then used to construct a dictionary of \qmo relevant volatility words\qmc, while the second part is used to perform backtesting. By \qmo relevant volatility words\qmc, we mean those words belonging to three different sentiment categories named \qmo GOOD\qmc, \qmo BAD\qmcsp and \qmo HIGH\qmc. In particular, accordingly to our taxonomy, \qmo GOOD\qmcsp, (\qmo BAD\qmc) volatility words are those that generate a positive (negative) market reaction, while \qmo HIGH\qmcsp are those that generate high volatility.
\noindent The analysis of the relevance of information to predict large losses starts by constructing a dictionary of \qmo relevant volatility words\qmc, which is subsequently used to deliver sentiment indicators for each sector we analyse. Sentiment indicators are then included as exogenous regressors in the models specified in Section \ref{sec:model}. To this end, the 2--minutes dataset is divided into two equal parts: the first part is used to build the dictionary while the second one is used to perform the out--of--sample analysis. By \qmo relevant volatility words\qmc, we mean those words belonging to three different sentiment categories named: \qmo Positive\qmc, \qmo Negative\qmcsp and \qmo High\qmc. In particular, accordingly to our taxonomy, \qmo Positive\qmc, or \qmo Negative\qmcsp volatility words are those that generate positive or negative market reactions, respectively. Those words generating high volatility reactions are instead classified as \qmo High\qmc.\newline
%%%
%The dictionary building up procedure starts from a raw classification of each in--sample headline. First we investigated for the \qmo GOOD/BAD\qmcsp response, i.e. headlines followed by a strong positive (negative) market movement and then we investigated for the \qmo HIGH\qmcsp response using the same criterium but considering the series of squared return as a proxy of the conditional variance.
%%%
\indent The dictionary building up procedure consists of two main steps. The first step classifies each in--sample headline into three different categories: \qmo Positive\qmc, \qmo Negative\qmcsp or  \qmo High\qmc. According to Koppel and Shtrimberg \citeyearpar{koppel_shtrimberg.2004} procedure, we search for which response the market had consequently to the headlines publishing. Headlines that are followed by a strong positive or negative market movement are classified as \qmo Positive\qmcsp or \qmo Negative\qmcsp news respectively, and those involving  subsequently large squared log--returns are classified as \qmo High\qmc. The evaluation of the market response is based on a two--minutes time frame instead of daily returns as suggested by Koppel and Shtrimberg \citeyearpar{koppel_shtrimberg.2004}.
%
%We instead differ from the Koppel and Shtrimberg \citeyearpar{koppel_shtrimberg.2004} classification procedure for the time span used to evaluate market responses because they used a daily time--frame while we account for the successive two minutes responses.
%
We decide to investigate the market news reaction using the successive two minute log--returns
because, as described in the Introduction, the semi strong form of the EMH states that stock prices will adjust to publicly available new information very rapidly. %The choice of daily cumulative returns of Koppel and Shtrimberg \citeyearpar{koppel_shtrimberg.2004} is instead motivated by the observation that financial markets are far from the semi strong form of the EMH.
While it is possible that financial markets are not aligned with the semi strong form of the EMH, the high number of spikes frequently observed by high frequency financial returns, suggests to use a short time frame. Regarding the choice of two minutes time span, we observe that the two--minutes return series have more accentuate spikes compared with the one, three and four minutes series.\newline
%
%We motivate this choice by considering the fact that, as described in the Introduction, the semi strong form of the EMH states that stock prices will adjust to publicly available new information very rapidly, so we decided to investigate the market news reaction using the successive two minute log--returns. The choice of Koppel and Shtrimberg \citeyearpar{koppel_shtrimberg.2004} is instead motivated by the observation that financial markets are far from the semi strong form of the EMH, and suggest to use the daily cumulative returns. While it is possible that financial markets are not aligned with the semi strong form of the EMH, the high number of spikes frequently observed by high frequency financial returns, suggests to use a short time frame. The choice of two minutes span seems to be somehow arbitrary, however, after several tests, it seems that the two minute return series has more accentuate spikes compared with the one, three and four minutes series.\newline
%
\indent Formally, we associate the two minutes log--return $r_t$ prior to the headline stamp, of the corresponding asset, to each in--sample headline $l_j$, for all $j=1,2,\dots,n_h$, where $n_h$ is the number of the in--sample headlines. Then, for given threshold levels, $\bar{r}_{\rm neg},\bar{r}_{\rm pos},\bar{r}_{\rm high}$, we associate the two classes $\mathcal{S}$=\{\qmo Positive\qmc, \qmo Negative\qmc, \qmo Neutral\qmc\}, and $\mathcal{Q}$=\{\qmo High\qmc, \qmo Low\qmc\} to each headline $l_j$ for $j=1,2,\dots,n_h$, according to the following rules:
\begin{itemize}
\item[-] $l_{j,\mathcal{S}}=$\qmo Negative\qmc, if $r_j \in \left(-\infty, \bar{r}_{\rm neg}\right]$, \qmo Positive\qmc,  if $r_j \in \left[\bar{r}_{\rm pos}, \infty\right) $, \qmo Neutral\qmc, if $r_j \in \left(\bar{r}_{\rm neg}, \bar{r}_{\rm pos}\right)$,
%\item $l_j$=\qmo Positive\qmcsp  if $r_j \in \left[r_{pos}, \infty\right) $\\
%\item $l_j$=\qmo Neutral\qmcsp  if $r_j \in \left(r_{neg}, r_{pos}\right] $\\
%
\item[-] $l_{j,\mathcal{Q}}=$\qmo High\qmc, if $r^2_j \in \left[\bar{r}_{\rm high}, \infty\right)$ and \qmo Low\qmc, if $r^2_j \in \left[0,\bar{r}_{\rm high}\right)$.
%\item $l_j$=\qmo Non--High\qmcsp  if $r^2_j \in \left[0, r_{high}\right) $\\
%
\end{itemize}
Threshold levels $\bar{r}_{\rm pos}$, $\bar{r}_{\rm neg}$ and $\bar{r}_{\rm high}$, are defined as the quantiles of the unconditional return distribution, i.e. $\mathbb{P}\left(r_t<\bar{r}_{\rm neg}\right) = 0.025$, $\mathbb{P}\left(r_t>\bar{r}_{\rm pos}\right) = 0.025$, and $\mathbb{P}\left(r_t^2<\bar{r}_{\rm high}\right) = 0.025$.\newline
%%%
\indent The second part of the regressor building procedure consists in categorising each word belonging to the news archive of 51266 headlines collected by FactSet for the in--sample period. As for the headlines classification, we first look for \qmo Positive\qmcsp and \qmo Negative\qmcsp volatility words, and then we search for \qmo High\qmcsp and \qmo Low\qmcsp volatility words. A preliminary classification was made by applying to each word the most frequent class of the headlines it belongs to. Formally, we assign to each word $w$ in every headline $l_j, j=1,2,\dots,n_h$ the modality of $\mathcal{S}$ and $\mathcal{Q}$ associated to the headlines in which each word $\omega$ appears most frequently. According to Das \citeyearpar{das_chen.2007}, in order to remain with only significant words, each word is subsequently scored using the Fisher discriminant statistic. Let $i$ be a class of the two categories $\mathcal{S}$=\{\qmo Positive\qmc, \qmo Negative\qmc, \qmo Neutral\qmc\}, and $\mathcal{Q}$=\{\qmo High\qmc, \qmo Low\qmc\}, and let $\mu_i$ be the average number of times the word $w$ appears in an headline of category $i$ and let $m_{i,j}$ be the number of times the word $w$ appears in a message $j$ of category $i$, then the Fisher discriminant statistic is defined as
\begin{eqnarray}
\mathrm{F}\left(w\right)
=\frac{\left(1/3\right)\sum_{i\neq k}\left(\mu_i-\mu_k\right)^2}{\left(\sum_{i}\sum_{j}\left(m_{i,j}-\mu_j\right)^2\right)/\left(\sum_{i}n_{i}\right)},
\label{eq:fisher_formula}
\end{eqnarray}
for any word $w$ previously categorised. The Fisher discriminant statistic delivers a score for each word which accounts for the level of category affiliation and we keep words with an acceptable level of $\mathrm{F}\left(w\right)$. At the end of this procedure we classified 259 words as \qmo Good\qmc, 239 as \qmo Bad\qmcsp and 366 as \qmo High\qmc. We remark that words which are \qmo Good\qmc, or \qmo Bad\qmcsp could also be \qmo High\qmcsp but words that are \qmo High\qmcsp are not necessary \qmo Good\qmcsp or \qmo Bad\qmc.\newline
%%%
\indent Based on the previous word classification, we built the following three regressors: \qmo POS\qmc, \qmo NEG\qmcsp and \qmo HIGH\qmc, measuring changes in the market sentiment. Each of the three regressors represents the number of words, of each categories, which appears in the interval $(t-1,t]$. It follows that, the inclusion of these regressors in the conditional variance dynamic allows it to change according to the new sentiment information coming from the news. We also considered two additional covariates: \qmo LAGVOL\qmcsp representing the lagged volumes and \qmo NUMB\qmcsp representing the total number of words in the interval $(t-1,t]$. While \qmo POS\qmc, \qmo NEG\qmcsp and \qmo HIGH\qmcsp are proxies for the market sentiment, \qmo NUMB\qmcsp and \qmo LAGVOL\qmcsp are a proxies for the information volume.
%%%
The regressor \qmo NUMB\qmcsp simply considers the total number of words as a proxy for the information arrival. It should be noticed that it does not coincide with the sum of the regressors \qmo POS\qmc, \qmo NEG\qmcsp and \qmo HIGH\qmcsp because even not categorised words are considered in the total number of words. The regressor \qmo NUMB\qmcsp accounts also for the news activism of a specific institution or sector. Indeed, it does not give any information about the nature of the news, but instead it simply considers the market interest into a specific institution or sector. The regressor \qmo LAGVOL\qmcsp instead is simply the aggregated lagged volumes at time $t$ and accounts for the market activism. In particular, it represents a measure of investors' interest into a specific institution or sector. Lagged volumes are also used in Kalev and Duong \citeyearpar{kalev_etal.2004}.
%%%
In order to use these regressors into the conditional volatility dynamics, the regressors' timestamp must be aligned with the 5--minute log--returns series. It follows that each regressor \qmo POS\qmc, \qmo NEG\qmc, \qmo HIGH\qmc, \qmo NUMB and \qmo LAGVOL\qmc, at time $t$ should account for the information arrival in the interval $\left(t-1,t\right]$ that generated the log--return a time $t$, $r_t$.\newline
%%%
\noindent For all the considered distribution errors and volatility specifications, we build the $\mathcal{M}_{\sI\sV}$ (Information Volumes) model which includes the variables \qmo NUMB\qmcsp and \qmo LAGVOL\qmcsp, and the $\mathcal{M}_{\sS\sE}$ (sentiment) model, which includes \qmo POS\qmcsp, \qmo NEG\qmcsp and \qmo HIGH\qmc covariates.
%
%we consider two different exogenous covariate specifications: the fist named \qmo Information Volumes\qmc, $\mathcal{M}_{\sI\sV}$, includes the variables \qmo NUMB\qmcsp and \qmo LAGVOL\qmcsp, and the second, named \qmo Sentiment\qmcsp $\mathcal{M}_{\sS\sE}$, includes \qmo POS\qmcsp, \qmo NEG\qmcsp and \qmo HIGH\qmc.
%
The $\mathcal{M}_{\sI\sV}$ specification represents the volume of the information flow, and can be considered as a measure of the investor interest in the institution or sector. Moreover, $\mathcal{M}_{\sI\sV}$ is a good candidate to improve the prediction of large market movements. The $\mathcal{M}_{\sS\sE}$ specification instead is a measure of sentiment and is useful to account for different market reactions to different kind of news. These two specifications are compared with the naive model $\mathcal{M}_{\sN}$ which does not consider any exogenous information.\newline
%%%
\indent Tables \ref{tab:IV_summary_stat} and \ref{tab:SE_summary_stat} report descriptive statistics for all the built regressors. The high level of kurtosis in the regressors' series (see for example the regressor \qmo NUMB\qmcsp in Table \ref{tab:IV_summary_stat}) could be linked with the high level of kurtosis usually reported by financial time series. In fact, according with the MDH hypothesis of Clark \citeyearpar{clark.1973}, the variance of returns is a function of the rate of information flow. More interesting, regressors exhibit even high level of autocorrelation, especially at lag 95 which coincides with one trading day at 5--minutes frequency. This phenomenon is probably due to the open--market imputation of overnight news. Indeed, in a high frequency timeframe, the news process has quite similar seasonal patterns as that of the volatility process: see e.g. Gencay \textit{et al.} \citeyearpar{gencay_etal.2001} and Kalev \textit{e.al} \citeyearpar{kalev_etal.2011} for a more comprehensive discussion.
%
%::::::::::::::::::::::::::::::::::::::::::::::::::::::::::::::::::::::
% SECTION: EMPIRICAL RESULTS
%::::::::::::::::::::::::::::::::::::::::::::::::::::::::::::::::::::::
\subsection{Empirical results}
\label{sec:empirical_results}
%::::::::::::::::::::::::::::::::::::::::::::::::::::::::::::::::::::::
%
\noindent In this section we discuss the results obtained by fitting the models to our dataset consisting of the 19 sectors of the STOXX Europe 600 index. For each sector, the models belonging to the initial set $\mathrm{M}^0$ obtained by combining the different volatility dynamics, the distributions for the error term and the covariates specifications discussed in Section \ref{sec:regressors}, are compared.\newline
%%
%For each sector, the initial set $\mathrm{M}^0$ of 27 models is obtained by combining the three GARCH--dynamics (GARCH, EGARCH, GJR--GARCH), the three errors distribution (Gaussian, Student--t and GED), and the three different exogenous covariate model specifications, $\mathcal{M}_{\sI\sV}$, $\mathcal{M}_{\sS\sE}$ and $\mathcal{M}_{\sN}$ discussed in Section \ref{sec:regressors}, are compared.\newline
%%%
% A/E RATIO
%%%
\indent A simple method to compare VaR forecasts is the actual versus expected rate (A/E), defined as the ratio between the realised VaR exceedances over a given time horizon and their \qmo a priori\qmcsp expected values. Table \ref{tab:A/E_all} reports the A/E ratio for all the ecompassing models at two VaR confidence levels $\tau=1\%$ and $\tau=0.1\%$. Table \ref{tab:A/E_all} shows that high levels of A/E for $\tau=0.1\%$ are usually associated to the Gaussian specification, reflecting its inadequacy to capture large price movements. Conversely, for the Student--t error distribution we observe levels of A/E closer to one. Moreover, for both the quantile confidence levels $\tau=\left(1\%,0.1\%\right)$, models accounting for the information volume $\left(\mathcal{M}_{\sI\sV}\right)$ or the market sentiment $\left(\mathcal{M}_{\sS\sE}\right)$ report value of the A/E ratio closer to one. A more deeper inspection reveals that, for $\tau=1\%$ the $\mathcal{M}_{\sN}$ model outperforms others only in 3 cases, while accounting for information volumes improves the VaR forecast ability in 12 cases. Moreover, as reported in Table \ref{tab:A/E_all}, even for $\tau=0.1\%$ the naive model $\mathcal{M}_{\sN}$ outperforms the specifications which include covariates in only two cases, while $\mathcal{M}_{\sI\sV}$ and $\mathcal{M}_{\sS\sE}$ report more accurate A/E levels 15 times over 19.\newline
% AD
\indent The magnitude of violating returns (AD) is also important, since it provides a measure of the expected loss given a VaR violation. Table \ref{tab:AD_all} reports the maximum AD of violating returns, as considered by McAleer and da Veiga \citeyearpar{mcaleer_daveiga.2008}. %for two confidence levels $\tau=\left(0.1\%, 1\%\right)$.
%%%
We can see that, for all the considered sectors, the $\mathcal{M}_{\sI\sV}$ and $\mathcal{M}_{\sS\sE}$ specifications exhibit lower maximum AD values with respect to %the model without exogenous information
the naive specification $\mathcal{M}_{\sN}$. This means that accounting for information helps to reduce the maximum AD loss at both high ($\tau=1\%$) and extreme ($\tau=0.1\%$) confidence levels. More interesting is to note how this result would be independent to the choice of the volatility dynamics or the error distribution specification. From a not reported analysis we found that the inclusion of the information flow also improves the general goodness--of--fit, according to both the AIC and BIC information criteria (those results are available upon request). For each of the six combinations of the GARCH--dynamics and the error distribution, the $\mathcal{M}_{\sI\sV}$ and $\mathcal{M}_{\sS\sE}$ specifications report lower levels of AIC and BIC.
% ACF
Moreover, the ACF of the standardised absolute residuals of the models $\mathcal{M}_{\sI\sV}$ and $\mathcal{M}_{\sS\sE}$ shows that the daily seasonal patterns are significantly reduced compared with the model $\mathcal{M}_{\sN}$. Once again accounting for the information flow helps to explain the volatility patterns which usually affects the high frequency financial time series.\newline
%
%In an unreported figure Figure \ref{fig:ACF_standardized_residuals} we report the ACF of the standardised absolute residuals of the models $\mathcal{M}_{\sI\sV}$ and $\mathcal{M}_{\sS\sE}$ where the daily seasonal patterns are significantly reduced compared with the model $\mathcal{M}_{\sN}$. Once again accounting for the information flow helps to explain the volatility patterns which usually affects the high frequency financial time series.\newline
%
%TO BE DONE: CC TEST, UC TEST AND DQ TEST\newline
%%%
\indent Tables \ref{tab:UC_test_kupiec}, \ref{tab:CC_test_christoffersen} and \ref{tab:DQ_test_engle_manganelli} instead report the unconditional coverage (UC) test of Kupiec \citeyearpar{kupiec.1995}, the conditional coverage (CC) test of Christoffersen \citeyearpar{christoffersen.1998} and the dynamic quantile (DQ) test of Engle and Manganelli \citeyearpar{engle_manganelli.2004}, at the 5\% confidence level. %for all the considered sectors for both the VaR series at $\tau=1\%$ and $\tau=0.1\%$.
The inclusion of news information does not report any discriminant result at the confidence level $\tau=1\%$. %In fact, all the computed tests for the confidence level $\tau=1\%$, report quite similar results for all the considered models.
Conversely, the results reported in tables \ref{tab:UC_test_kupiec}, \ref{tab:CC_test_christoffersen} and \ref{tab:DQ_test_engle_manganelli} for the extreme VaR confidence level $\tau=0.1\%$ confirm that the choice of a heavy tailed distribution for the error term, is strongly discriminant. Furthermore, for $\tau=0.1\%$, the rejection of the null hypothesis of both tests is more frequent for the naive specification $\mathcal{M}_{\sN}$.\newline
%
%It is worth noting that for the built dataset we also test the Skewed-GARCH counterpart of the assumed distribution (see for example De Luca and Loperfido \citeyearpar{deluca_loperfido.2004}). In this case the estimated shape parameters equal to one indicate the absence of skewness in the residuals.
%%%
%%
%Tables summarising the in--the--sample parameter estimates for all the competing model specifications are available as supplementary material. For some sectors, the parameters associated to the conditional mean dynamic appear to be not statistically significant. This essentially means that for those series the well know stylised fact of negative serial autocorrelation frequently displayed by high frequency data is not present. As described by Gencay \textit{et al.} \citeyearpar{gencay_etal.2001}, this phenomenon can be associated with the procedure employed to construct the high frequency series. Even parameters associated with the exogenous regressor sometimes appears to be not statistically significant, but this does not represent a problem for our analysis since it is mainly focused on extreme loss prediction and not on describing the whole returns behaviour. %However, we do not really care about that since our analysis is mainly focused on extreme loss prediction and not on describing the whole returns behaviour.
%%
\indent From an unreported in--the--sample parameter estimation analysis (available to the authors upon request) we found that for some sectors, the parameters associated to the conditional mean dynamic appear to be not statistically significant. This essentially means that for those series the well know stylised fact of negative serial autocorrelation frequently displayed by high frequency data is not present. As described by Gencay \textit{et al.} \citeyearpar{gencay_etal.2001}, this phenomenon can be associated to the procedure employed to construct the high frequency series. In addiction, our results showed that some parameters associated with the exogenous regressors appear to be not statistically significant which does not represent a problem for this study since our focus is on extreme loss prediction and not on describing the whole returns behaviour.\newline %However, we do not really care about that since our analysis is mainly focused on extreme loss prediction and not on describing the whole returns behaviour.
%%%
\indent Concerning the specification of the error distribution, we also test for the presence of asymmetry by estimating the Skew--GARCH specifications already introduced by De Luca and Loperfido \citeyearpar{deluca_loperfido.2004}. In particular, we considered the skewing mechanism of Fernandez and Steel \citeyearpar{fernandez_steel.1998} which guarantees a regular behaviour of the likelihood function. The estimated shape parameters are equal to one for all the considered sectors, indicating that the skewed distributions collapse to the symmetric counterparts.\newline
%
%De Luca G., Loperfido N. (2004) A skew-in-mean GARCH model for financial returns, in: Skew-Elliptical Distributions and Their Applications: a Journey Beyond Normality, Genton M.G. (Ed.), CRC/Chapman & Hall, 205-222.
%
\indent In next section we report results for the Model Confidence Set procedure.
%
%::::::::::::::::::::::::::::::::::::::::::::::::::::::::::::::::::::::
% SECTION: MCS RESULTS
%::::::::::::::::::::::::::::::::::::::::::::::::::::::::::::::::::::::
\subsection{Application of the MCS procedure}
\label{sec:MCS_results}
%::::::::::::::::::::::::::::::::::::::::::::::::::::::::::::::::::::::
%
\noindent In this section we discuss the results we obtained applying the MCS procedure of Hansen \citeyearpar{hansen.2003} and Hansen \textit{et al.} \citeyearpar{hansen_etal.2011}.
%%%
In Table \ref{tab:MCS_composition_final} we report, for each sector and each covariate specification, the number of models belonging to the SSM at the end of the MCS procedure, i.e. $\hat{\rm M}_{1-\alpha}^*$. Clearly, a small final set of models means that the procedure has been highly discriminant, i.e. models that performs poorly were eliminated. Conversely, a large final set of models means that there were not statistical evidence of different performances among the competing models.
%%%
As shown in Table \ref{tab:MCS_composition_final} the MCS procedure reports highly discriminant results for most of the sectors and for both the VaR confidence levels $\tau=\left(1\%,0.1\%\right)$. Comparing the results for the two confidence levels we observe that the final set contains only 7 occurrences for $\tau=0.1\%$, while even 16 for $\tau=1\%$. This means that, using the MCS procedure, the set of models that adequately predicts extreme VaR levels, such those implied by a confidence level of 0.1\% is lower than that predict higher confidence levels VaR. This result is in line with the idea that only few models are able to predict large losses.
Moreover, it is clear from table \ref{tab:MCS_composition_final} that, performing the MCS procedure, the model that most appears in the final set $\hat{\rm M}_{1-\alpha}^*$ is $\mathcal{M}_{\sI\sV}$. This means that adding information volume proxies in the volatility dynamics, improve models ability to predict large losses.\newline
%%%
\indent Table \ref{tab:MCS_average_AE_AD} reports, for each covariate specification, the Actual over Expected (A/E) exceedance ratio and the maximum AD violating returns averaged across the models belonging to the optimal final set $\hat{\rm M}_{1-\alpha}^*$ delivered by the MCS procedure. We observe that, for those cases where the MCS procedure does not discriminate models, the average A/E ratio reports inconclusive results, while the averaged maximum AD violating loss shows that the specification $\mathcal{M}_{\sI\sV}$ is preferred for most of the considered sectors. Finally, we observe that the specification $\mathcal{M}_{\sI\sV}$ corresponds always to the best performing model under the AD metric for all the considered sectors.
%
%::::::::::::::::::::::::::::::::::::::::::::::::::::::::::::::::::::::
% SECTION: VAR COMBINATION RESULTS
%::::::::::::::::::::::::::::::::::::::::::::::::::::::::::::::::::::::
\subsection{VaR combination results}
\label{sec:VaR_combination_results}
\begin{figure}[h]
\centering
\captionsetup{width=1.0\textwidth}
\includegraphics[width=1.0\linewidth, height=0.66\linewidth]{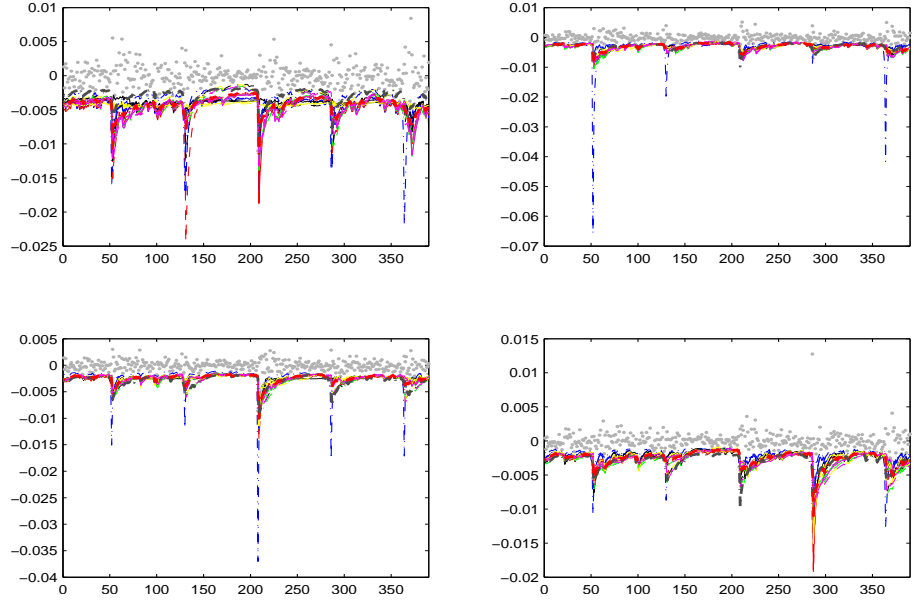}
\caption{\footnotesize{Time series plot of the log--returns \textit{(grey bullets)} along with the estimated VaR at $\tau=1\%$ for different sectors: communications \textit{(upper, left)}, utilities \textit{(upper, right)}, distribution services \textit{(bottom, left)}, health technology \textit{(bottom, right)}. In all panels, thin lines denote VaR forecast delivered by individual models ${\rm VaR}_{j,t+1\vert t}^{\tau}$, $\forall j=1,2,\dots,\mathsf{M}_0$, while the grey dotted line denotes the ${\rm VaR}_{t+1\vert t}^{\tau,{\rm avg}}$ and the red dotted line denotes the ${\rm VaR}_{t+1\vert t}^{\tau,{\rm dyn}}$ obtained by averaging the model specifications belonging to the MCS set $\mathsf{M}_{1-\alpha}^*$.}}
\label{fig:VaR_ALL}
\end{figure}

\begin{figure}[ht]
\centering
\captionsetup{width=1.0\textwidth}
\includegraphics[width=1.0\linewidth, height=0.66\linewidth]{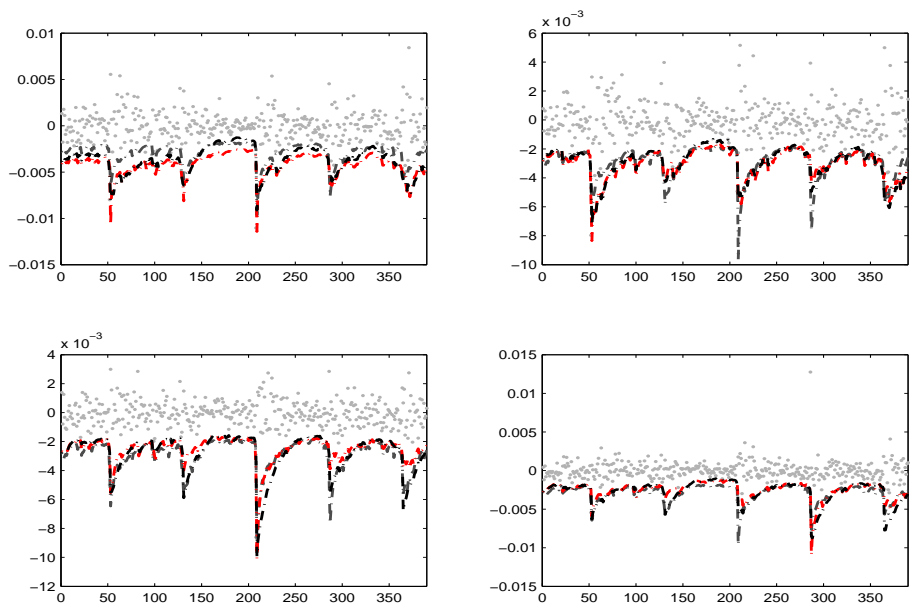}
\caption{%
\footnotesize{Time series plot of the log--returns \textit{(grey bullets)} along with the estimated VaR at $\tau=1\%$ for different sectors: communications \textit{(upper, left)}, utilities \textit{(upper, right)}, distribution services \textit{(bottom, left)}, health technology \textit{(bottom, right)}. In all panels the grey dotted line denotes the ${\rm VaR}_{t+1\vert t}^{\tau,{\rm avg}}$, the red dotted line denotes the ${\rm VaR}_{t+1\vert t}^{\tau,{\rm dyn}}$ obtained by averaging all the model specifications in the initial set $\mathsf{M}_0$, and the dark dotted line denotes the ${\rm VaR}_{t+1\vert t}^{\tau,{\rm dyn}}$ obtained by averaging the model specifications belonging to the MCS set $\mathsf{M}_{1-\alpha}^*$.}}
\label{fig:VaR_Dyn_Avg}
\end{figure}

%::::::::::::::::::::::::::::::::::::::::::::::::::::::::::::::::::::::
%
\noindent The main aim of the VaR combination technique introduced in Section \ref{sec:combining_vars} is to provide a practical tool that combines VaR forecasts at a given quantile level $\tau$ delivered by different models. The MCS procedure ensures the availability of a set of model having the same ability to predict the VaR at the desired quantile level, under the chosen loss function. The VaR forecast delivered by those models can subsequently be used as the main ingredients of the VaR combination technique previously detailed.\newline
%%%
%
%I denote the autoregressive coefficient of model $k$ belonging to sector $j$ at the VaR confidence level $\tau$ by $\kappa_{k,j}^\tau$, where $k=1,\dots,\#\left(j,\tau\right)$ and $\#\left(j,\tau\right)$ is a function the counts the number of models in the SSM of sector $j$ at confidence level $\tau$. Obviously when $\#\left(j,\tau\right)=1$ means that only one model is present in the SSM and consequently the VaR combination technique is not needed. Following the same notation it is possible to denote by $\omega_{k,j,t}^\tau$ the weight of model $k$ belonging to the sector $j$ at time $t$.\\
%
%\indent Tables \ref{tab:kappa_all_1} and \ref{tab:kappa_all_0.1} reports the estimated $\hat\kappa_{k,j}^\tau\quad j=1,\dots,19$ for all models when $\tau=1\%$ and $\tau=0.1\%$. When $\hat\kappa_{k,j}^\tau$ is small, the weight's dynamic is not so smooth, moreover the impact to the overall weights $\hat\omega_{k,j,t}^\tau,\quad k=1,\dots,\#\left(j,\tau\right)$ of the changes in the contribution of model $k$'s loss into the average loss\footnote{This is the quantity $\chi_{u,t}$ presented in Section \ref{sec:new_model}, that in this contest can be indicated with $\chi_{k,j,t}^\tau$ following the previous notations} is high. Conversely when $\hat\kappa_{k,j}^\tau$ is large (say > 0.8) the weight dynamic is quite smooth over time and the relative loss presented in Equation \eqref{eq:mydynamic_parts} does not have an high contribution to the overall weights.\\
%%%
\indent In order to test the benefits of using the proposed technique and the MCS procedure we report a simple comparison using two different VaR forecasts combinations. The first VaR forecast simply averages the individual VaRs across the models belonging to the MCS
\begin{equation}
{\rm VaR}^{\tau,{\rm avg}}_{t+1\vert t}=\frac{1}{m^*}\sum_{j=1}^{m^*}{\rm VaR}^\tau_{j,t+1\vert t},
\end{equation}
%
%\frac{\sum_{k=1}^{m}\mathrm{VaR_{k,j,t}^\tau}}{m}
for $t=1,2,\dots,T$. The second model VaR forecast results by applying the VaR combination technique proposed in equations \eqref{eq:var_combination} and \eqref{eq:vars_weighing_equation}. Table \ref{tab:var_comb_comparison}, reports the mean and the maximum AD for the two forecasting combination methods and the two VaR confidence levels $\tau=\left(1\%,0.1\%\right)$. Inspecting Table \ref{tab:var_comb_comparison} it is possible to see the benefit of using the MCS procedure along with the proposed VaR combination technique instead of simply averaging the available models. For all the considered sectors and for both the confidence levels, the dynamic VaR combination approach ${\rm VaR}_{t+1\vert t}^{\tau,{\rm dyn}}$ reports values of the max AD being almost the 50\% lower than those reported by the simple VaR average ${\rm VaR}_{t+1\vert t}^{\tau,{\rm dyn}}$. This suggests that the proposed VaR combination technique helps in reducing the large and extreme absolute deviations of VaR violating returns. Looking at columns 6 to 9 of Table \ref{tab:var_comb_comparison} it is possible to note that the proposed technique helps also in reducing the mean AD of VaR violating returns which is particularly relevant from a risk management point of view, since reducing the mean AD corresponds to a lower expected loss when a VaR violation occurs. In Figures \ref{fig:VaR_ALL} and \ref{fig:VaR_Dyn_Avg} we report an illustrative comparison between the VaRs delivered by individual model ${\rm VaR}_{j,t+1\vert t}^{\tau}$, for $j=1,2,\dots,m^{*}$, the VaR forecast obtained by the static average ${\rm VaR}_{t+1\vert t}^{\tau,{\rm avg}}$ and the dynamic VaR forecasts ${\rm VaR}_{t+1\vert t}^{\tau,{\rm din}}$ series.\newline
%%%
\indent As previously detailed, the VaR combination technique provides a dynamic linear convex combination of the available VaR series; this is showed by Figure \ref{fig:VaR_ALL} and \ref{fig:VaR_Dyn_Avg}, where several VaR forecast series along with the ${\rm VaR}_{t+1\vert t}^{\tau,{\rm avg}}$ \textit{(grey dotted line)} and the ${\rm VaR}_{t+1\vert t}^{\tau,{\rm dyn}}$ \textit{(red dotted line)} are reported.
%%%
Since the autoregressive parameters of the dynamic VaR combination are estimated by minimizing the asymmetric loss function of Gonzalez--Riviera \textit{et al.} \citeyearpar{gonzalez_etal.2004}, it follows that the resulting combined VaR series is more robust against outliers than the static ${\rm VaR}_{t+1\vert t}^{\tau,{\rm avg}}$ series. We note for example that models which are highly non linear and strongly sensitive to outliers, such as the EGARCH (thin blue line, in Figure \ref{fig:VaR_ALL}), sometimes results in wrong VaR predictions. However, since they can be useful to describe others complex financial dynamics of the conditional variance process it is important to use them in the dynamic VaR averaging. Our proposed VaR combination technique results therefore in a practical tool to account for several models preventing the outlier consequences on VaR forecasts. Moreover, Figure \ref{fig:VaR_Dyn_Avg} compares the static VaR combination \textit{(grey dotted line)} with those obtained by using our dynamic technique in the cases where all the models belonging to the initial set $\mathsf{M}_0$ \textit{(red dotted line)} or only the model belonging to the MCS final set $\mathsf{M}_{1-\alpha}^*$ \textit{(dark dotted line)} are considered. Figure \ref{fig:VaR_Dyn_Avg} reveals that the dynamic method combined with the MCS procedure \textit{(dark dotted line)} provides more conservative VaR estimate especially in presence of extreme drop-downs.
%
% there are some highly non linear models, such as the EGARCH (thin blue dotted), for which an outlier usually result in a wrong VaR prediction; from a risk management point of view this means more capital to hold for regulatory requirement. However, those highly non linear models can be useful to describe others complex dynamics of the conditional variance process; the proposed VaR combination technique provides a practical tool in order to account for several models preventing the outlier consequences on VaR forecasts.
%
%::::::::::::::::::::::::::::::::::::::::::::::::::::::::::::::::::::::
% SECTION: CONCLUSION
%::::::::::::::::::::::::::::::::::::::::::::::::::::::::::::::::::::::
\section{Conclusion}
\label{sec:conclusion}
%::::::::::::::::::::::::::::::::::::::::::::::::::::::::::::::::::::::
%
\noindent In this study we investigate whether the inclusion of quantitative measures of information news could improve volatility models to predict the extreme financial returns frequently observed in financial time series sampled at high frequencies. To address this problem, we construct three basic sentiment indexes to measure the impact of \qmo positive\qmc, \qmo negative\qmc and \qmo high\qmcsp volatility words, starting from the raw information flow provided by \qmo FactSet\qmc. The study considers all the STOXX Europe 600 index constituents sampled at high--frequency, subsequently aggregated in sectors according to the \qmo FactSet\qmcsp classification. For each sector, we consider 27 competing models obtained by combining three GARCH- dynamics specifications (GARCH, EGARCH, GJR--GARCH), three error distributions (Gaussian, Student--t and GED), and three exogenous covariates models: $\mathcal{M}_{\sI\sV}$ accounting for the information volume, $\mathcal{M}_{\sS\sE}$ accounting for sentiment indexes and $\mathcal{M}_{\sN}$ where no exogenous regressors are considered.
%%%
The predictive ability of each model specification is then tested using several criteria.
%%%
We observe that the inclusion of information volumes $\mathcal{M}_{\sI\sV}$ improve the VaR forecast ability, the general goodness--of--fit, and reduce the maximum loss after a VaR violation compared to the naive model $\mathcal{M}_{\sN}$. Statistics for the A/E ratio, AIC and BIC information criteria (not reported to save space), and maximin AD losses are considered for each sector at both VaR confidence levels $\tau=1\%$ and $\tau=0.1\%$. We also observe that the introduction of exogenous regressors accounting for the information flow, helps also to explain the seasonal volatility patterns witch are present in financial data sampled at high frequency.
%%%
Moreover, since different models are available to describe the same data generating process, we adapt the sequential testing procedure MSC of Hansen \citeyearpar{hansen_etal.2011} to deliver a set of statistically equivalent models in terms of their ability to forecast one--step--ahead VaRs. To achieve this goal we propose to use a loss function penalising in an asymmetric way returns above and below the forecasted VaR proposed by Gonzalez--Riviera \textit{et al.} \citeyearpar{gonzalez_etal.2004} for backtesting purposes.
%
%In order to forecast ability under the chosen loss function during a pre specified evaluation period. Our out--of--sample evaluation period includes 4357 one step ahead rolling forecasts. The specified quantile loss function of Gonzalez--Riviera \textit{et al.} \citeyearpar{gonzalez_etal.2004}, provides a way to measure and compare deviations from the predicted VaRs.
%
%for two VaR confidence levels: large $\tau=1\%$ and extreme $\tau=0.1\%$.
%
%Our out--of--sample evaluation period includes 4357 one step ahead rolling forecasts. The Model Confidence Set procedure delivers a set of statistically equivalent models in terms of their one--step--ahead forecast ability under the chosen loss function.\newline
%%%
%\indent Concerning the results we obtain from our empirical investigation, we observe that the inclusion of information volumes $\mathcal{M}_{\sI\sV}$ improve the VaR forecast ability, the general goodness--of--fit, and reduce the maximum loss after a VaR violation compared to the naive model $\mathcal{M}_{\sN}$. Statistics for the A/E ratio, AIC and BIC information criteria, and maximin AD losses are reported for each sector at both VaR confidence levels. Moreover, we observe that the introduction of exogenous regressors accounting for the information flow, helps also to explain the seasonal volatility patterns witch are present in the high frequency financial data.
%
%Our additional findings regard the application of the MCS procedure.
%
Our results confirms that the MCS provides evidence in favour of $\mathcal{M}_{\mathsf{IV}}$ models showing that adding information volume proxies in the volatility dynamics, improves models ability to predict large losses. Unfortunately, in particular for small confidence levels, we can not uniquely conclude for the superior VaR forecasting performances of the models including information, because the MCS final set does not contains only $\mathcal{M}_{\mathsf{IV}}$ models. For those circumstances we propose a new VaR estimator that combines the VaRs delivered by those models belonging to the SSM using a smoother moving average process that guarantees an asymmetric penalisation of returns above and below the forecasted VaRs which account for the magnitude of VaR violations. We show that the MCS procedure along with the VaR combination technique results in a reduced mean and maximum AD of VaR violations suggesting that the VaR combination technique helps to reduce the magnitude of large and extreme VaR violations.\newline %The reduction of the mean AD of violations means a lower expected loss when a VaR violation occurs.
%
%In particular, while for those sectors where the final set is composed only by models accounting for the information volume, we have strong evidence of $\mathcal{M}_{\mathsf{IV}}$'s superior ability, and we can uniquely conclude for the superior VaR forecasting performances of the models including information, the same is not true for the sectors where the final MCS set is made of different kind of models. For those final sets if we consider the best model as the one with the A/E ratio closest to one we can conclude that the Naive models never outperforms models accounting for news.\newline
%%%
%\indent However, these results can not be considered conclusive when different models provide the same forecasting VaR performances and the MCS procedure is not able to discriminate them. In those circumstances a new VaR estimator that combines the VaR estimates delivered by those models belonging to the SSM. The proposed VaR estimator is simply a convex linear combination with time--varying weights. The proposed technique is particularly useful when more VaR forecasts are available and it is not obvious which one is the best. Our results show that the MCS procedure along with the VaR combination technique really results in a reduced mean and maximum AD of VaR violations suggesting that the VaR combination technique helps to reduce the magnitude of large and extreme VaR violations. The reduction of the mean AD of violations means a lower expected loss when a VaR violation occurs.
\indent The analysis carried out throughout  this paper can be further extended to account for additional proxies measuring different news affecting the stock market. In particular, general macroeconomic news can be distinguished by those affecting only individual institutions. Moreover, it may be interesting to investigate the impact of information during different periods of market turbulence.

%
%:::::::::::::::::::::::::::::::::::::::::::::::::::::::::::::::::::::
% SUBSECTION: ACKNOWLEDGMENTS
%:::::::::::::::::::::::::::::::::::::::::::::::::::::::::::::::::::::
\section*{Acknowledgments}
%:::::::::::::::::::::::::::::::::::::::::::::::::::::::::::::::::::::
%
\noindent This research is supported by the Italian Ministry of Research PRIN 2013--2015, ``Multivariate Statistical Methods for Risk Assessment'' (MISURA), and by the ``Carlo Giannini Research Fellowship'', the ``Centro Interuniversitario di Econometria'' (CIdE) and ``UniCredit Foundation''. A special thank goes to Riccardo Sucapane for his constructive comments and helps in particular to parallelise the R--code.
%%%
%
%::::::::::::::::::::::::::::::::::::::::::::::::::::::::::::::::::::::::::
% Section: Appendix A
%::::::::::::::::::::::::::::::::::::::::::::::::::::::::::::::::::::::::::

\appendix
%::::::::::::::::::::::::::::::::::::::::::::::::::::::::::::::::::::::
\section{Bootstrap resampling}
%::::::::::::::::::::::::::::::::::::::::::::::::::::::::::::::::::::::
\label{sec:appendix_A}
%::::::::::::::::::::::::::::::::::::::::::::::::::::::::::::::::::::::
%
\noindent In this appendix we detail the algorithm used to perform the block--bootstrap to obtain the bootstrapped estimates of ${var}\left(\bar d_{ij}\right)$ and the p--values of the MCS statistics of Hansen \citeyearpar{hansen_etal.2011} described in Section \ref{sec:model_sel}. The bootstrap procedure returns $B$ clones that mimic the structure of the original data. In order to do this, we first estimate the block--length $k$, using the maximum number of significant parameters of an AR(p) fitted on the original series; obviously a different criteria can be used, or different lengths can be tested to verify that results is not too sensitive to the chosen length $k$. The procedure can be summarised as follow.
\begin{itemize}
\item[1.] Set the maximum number of block $v=\lfloor n/k\rfloor-1$, where $\lfloor a\rfloor$ denotes the smallest integer greater or equal to $a$.
\item[2.] For each $j=1,2,\dots,v$, extract $\xi_{j}^{b}\sim \mathcal{U}\left(1,\dots,n\right)$, where $\mathcal{U}$ is a discrete uniform distribution on $\left\{1,2,\dots,n\right\}$ and the vector $\bxi^{b}=\left(\xi_1^b,\xi_2^b,\dots,\xi_v^b\right)$ of length $v$ contains the starting index of each block.
\item[3.] Construct the $\left(k\times v\right)$ matrix $\bXi^b$ in such a way that the first row contains the vector $\bxi^b$ and subsequent elements are determined as follows: set $\Xi_{i,j}^{b}=\Xi_{i-1,j}^{b}+1$ for $i=2,\dots,k$ and $j=1,\dots,v$, using the convention that $\Xi_{i,j}^{b}=\Xi_{i,j}^{b}-n$ for those $\Xi_{i,j}^{b}>n$.

\item[5.] Concatenate each column of $\bXi^b$ in a vector $\bb^{b}=\vec\left(\bXi^b\right)$ of length $\left(kv\times 1\right)$ and if $l=n-kv>0$ add $l$ new indexes extracting from $\mathcal{U}\left(1,\dots,n\right)$.

\item[6.] Extract, from the original series, the observation having the indexes contained in $\bb^{b}$ and store it.

\item[7.] Repeat steps 2--6, for $b=2,3,\dots,B$.
\end{itemize}
Using the $B$ bootstrap samples obtained from the bootstrap procedure previously described, it is possible to estimate ${\rm var}\left(\bar d_{i\cdot}\right)$  and ${\rm var}\left(\bar d_{ij}\right)$ that are necessary to construct the statistic tests, in the following way:
\begin{itemize}
\item[1.] define the bootstrap re--samples averages for $\bar{d}_{ij}$ as follows
\begin{equation}
\bar d_{ij}^{b}=\frac{1}{n}\sum_{t=1}^n d_{ij,t}^b\nonumber
\end{equation}
\item[2.] estimate the bootstrapped variance as
\begin{equation}
\widehat{\rm var}\left(\bar{d}_{ij}\right)=\frac{1}{B}\sum_{b=1}^{B}\left(\bar{d}_{ij}^{b}-\bar{d}_{ij} \right).\nonumber
\end{equation}
\end{itemize}
Obviously, the asymptotic distribution of the statistic test $\mathrm{T}_{\max}$ under the null hypothesis of the model confidence set test defined in Section \ref{sec:model_sel}, is nonstandard and it could be estimated using the empirical cdfs obtained from the previous bootstrap replications:
\begin{eqnarray}
\mathrm{T}_{\rm R,M}^{b}=\max_{i,j \in M} \frac{\mid \bar{d}_{ij}^{b}-\bar{d}_{ij} \mid}{\sqrt{\widehat{\rm var}\left(\bar{d}_{ij}\right)}}.\nonumber
\end{eqnarray}
Finally, the p--values could be calculated as:
\begin{eqnarray}
\mathrm{P}_{\rm R,M}=\frac{1}{B}\sum_{b=1}^B \bbone_{\mathrm{T}_{\rm R,M} >\mathrm{T}_{\rm R,M}^{b}}.\nonumber
\end{eqnarray}
%
%::::::::::::::::::::::::::::::::::::::::::::::::::::::::::::::::::::::::::
% Section: Appendix B
%::::::::::::::::::::::::::::::::::::::::::::::::::::::::::::::::::::::::::
\newpage
\clearpage
%
%::::::::::::::::::::::::::::::::::::::::::::::::::::::::::::::::::::::
\section{Tables}
%::::::::::::::::::::::::::::::::::::::::::::::::::::::::::::::::::::::
%\label{sec:appendix_B}
%::::::::::::::::::::::::::::::::::::::::::::::::::::::::::::::::::::::
%
%::::::::::::::::::::::::::::::::::::::::::::::::::::::::::::::::::::::
% TABLE: MCS PERCENTAGE COMPOSITION
%::::::::::::::::::::::::::::::::::::::::::::::::::::::::::::::::::::::
\begin{table}[!h]
\centering
\begin{tabular}{lccccccccc}
\toprule
 $\tau$& \multicolumn{4}{c}{$1\%$} & \multicolumn{4}{c}{$0.1\%$}\\
 \cmidrule(lr){1-1}\cmidrule(lr){2-5}\cmidrule(lr){6-9}
 Sector & $\mathcal{M}_{\mathsf{IV}}$ & $\mathcal{M}_{\mathsf{SE}}$ & $\mathcal{M}_{\mathsf{N}}$ & \# & $\mathcal{M}_{\mathsf{IV}}$ & $\mathcal{M}_{\mathsf{SE}}$ & $\mathcal{M}_{\mathsf{N}}$ & \# \\
 \cmidrule(lr){1-1}\cmidrule(lr){2-5}\cmidrule(lr){6-9}
Commercial services & 100 & 0 & 0 & 1 & 100 & 0 & 0 & 4 \\
Communications & 100 & 0 & 0 & 6 & 100 & 0 & 0 & 1 \\
Consumer durables & 100 & 0 & 0 & 1 & 100 & 0 & 0 & 5 \\
Consumer non-durables & 25 & 38 & 38 & 24 & 33 & 33 & 33 & 18 \\
Consumer services & 32 & 32 & 37 & 19 & 35 & 29 & 35 & 17 \\
Distribution services & 100 & 0 & 0 & 2 & 100 & 0 & 0 & 6 \\
Electronic Technology & 26 & 39 & 35 & 23 & 38 & 31 & 31 & 16 \\
Energy Minerals & 100 & 0 & 0 & 1 & 32 & 41 & 27 & 22 \\
Finance & 0 & 18 & 82 & 11 & 33 & 33 & 33 & 18 \\
Health Services & 100 & 0 & 0 & 1 & 33 & 33 & 33 & 18 \\
Health Technology & 0 & 36 & 64 & 11 & 33 & 28 & 39 & 18 \\
Industrial Services & 100 & 0 & 0 & 4 & 30 & 35 & 35 & 20 \\
Non-Energy Minerals & 100 & 0 & 0 & 1 & 100 & 0 & 0 & 5 \\
Process Industries & 100 & 0 & 0 & 1 & 35 & 30 & 35 & 23 \\
Producer Manufacturing & 100 & 0 & 0 & 5 & 100 & 0 & 0 & 1 \\
Retail Trade & 100 & 0 & 0 & 1 & 37 & 32 & 32 & 19 \\
Technology Services & 25 & 38 & 38 & 24 & 38 & 25 & 38 & 16 \\
Transportation & 100 & 0 & 0 & 5 & 40 & 27 & 33 & 15 \\
Utilities & 100 & 0 & 0 & 6 & 42 & 26 & 32 & 19 \\
\bottomrule
\end{tabular}
\caption{\footnotesize{Total number and percentage composition of models belonging to the SSM, $\hat{\rm M}_{1-\alpha}^*$, at the end of the MCS procedure, for each sector and covariate specification. We consider two confidence levels $\tau=0.1\%$ and $\tau=1\%$.}}
\label{tab:MCS_composition_final}
\end{table}
%
%::::::::::::::::::::::::::::::::::::::::::::::::::::::::::::::::::::::
% TABLE: DATA SUMMARY STATISTICS of RETURNS
%::::::::::::::::::::::::::::::::::::::::::::::::::::::::::::::::::::::
\begin{sidewaystable}[!h]
\captionsetup{font={small}, labelfont=sc}
\begin{center}
\begin{small}
\resizebox{0.8\columnwidth}{!}{%
 \smallskip
  \begin{tabular}{lcccccccc}\\
  \hline
   Name & Min & Max & Mean$\times10^3$ & Std. Dev. & Skewness & Kurtosis & 1\% Str. Lev. & JB\\
    \hline
   %  \multicolumn{9}{c}{\textit{International Stock Indeces}}\\
Electronic Technology & -1.401 & 1.892 & 0.324 & 0.107 & 0.712 & 34.187 & -0.291 & 425303.609  \\
Non-Energy Minerals & -2.316 & 2.457 & 0.336 & 0.174 & 0.372 & 27.24 & -0.431 & 269749.330  \\
Health Technology & -0.959 & 1.103 & 1.436 & 0.073 & 0.431 & 23.514 & -0.188 & 201117.210  \\
Distribution services & -1.549 & 1.203 & 1.182 & 0.077 & -0.203 & 42.982 & -0.184 & 671177.482  \\
Finance & -1.004 & 1.183 & 1.580 & 0.087 & 0.365 & 21.263 & -0.218 & 164436.848  \\
Technology Services & -1.415 & 1.510 & 0.675 & 0.094 & -0.542 & 36.129 & -0.234 & 474597.978  \\
Process Industries & -1.044 & 1.271 & 1.378 & 0.087 & 0.221 & 20.02 & -0.223 & 145674.251  \\
Commercial services & -1.058 & 1.382 & 1.624 & 0.084 & 0.883 & 35.355 & -0.205 & 455194.806  \\
Utilities & -0.555 & 0.762 & 0.019 & 0.073 & 0.215 & 10.556 & -0.207 & 40552.287  \\
Health Services & -1.088 & 1.746 & 1.609 & 0.083 & 1.269 & 46.001 & -0.201 & 771047.391  \\
Retail Trade & -0.647 & 1.275 & 0.607 & 0.082 & 0.638 & 17.235 & -0.220 & 108503.601  \\
Energy Minerals & -1.258 & 1.094 & 1.296 & 0.123 & 0.190 & 12.471 & -0.309 & 56553.002  \\
Consumer services & -1.097 & 1.546 & 1.601 & 0.091 & 1.611 & 39.201 & -0.218 & 562015.749  \\
Consumer non-durables & -0.778 & 0.827 & 0.747 & 0.063 & 0.728 & 22.188 & -0.162 & 179610.301  \\
Producer Manufacturing & -1.204 & 2.031 & 1.546 & 0.100 & 1.715 & 46.850 & -0.244 & 801616.206  \\
Industrial Services & -1.351 & 1.573 & 0.905 & 0.108 & 0.291 & 19.142 & -0.278 & 133230.915  \\
Communications & -1.967 & 3.856 & 0.436 & 0.117 & 3.00 & 161.957 & -0.282 & 9541186.331  \\
Transportation & -1.288 & 1.269 & 0.864 & 0.094 & -0.198 & 27.070 & -0.235 & 266254.814  \\
Consumer durables & -1.126 & 3.027 & 2.192 & 0.111 & 2.696 & 73.858 & -0.279 & 1992147.898  \\
\hline
\end{tabular}}
\caption{\footnotesize{Summary statistics of each sector log--returns in percentage points, for the period 17/08/2012 09:00:00 to 01/02/2013 09:40:00. The seventh column, denoted by ``1\% Str. Lev.'' is the 1\% empirical quantile of the returns distribution, while the last column, denoted by \qmo JB\qmcsp is the value of the Jarque-Ber\'a test--statistics.}}
\label{tab:Sectors_return_summary_stat}
\end{small}
\end{center}
%
%mallskip
%
\end{sidewaystable}
%
%
%::::::::::::::::::::::::::::::::::::::::::::::::::::::::::::::::::::::
% TABLE: DATA SUMMARY STATISTICS of
% EXOGENOUS REGRESSORS IV
%::::::::::::::::::::::::::::::::::::::::::::::::::::::::::::::::::::::
\begin{sidewaystable}[!ht]
\captionsetup{font={small}, labelfont=sc}
%\begin{table}[ht]
\centering
 \smallskip
  \centering
   \resizebox{1.0\columnwidth}{!}{%
    \setlength\tabcolsep{1mm}
 \begin{tabular}{lccccccccccccc}\\
  \hline
 & \multicolumn{6}{c}{LAGVOL} & \multicolumn{6}{c}{NUMB} \\
\cmidrule(lr){1-1}\cmidrule(lr){2-7}\cmidrule(lr){8-13}
   Name & Max & Mean & Std. Dev. & Skewness & Kurtosis & JB$\times10^{-6}$ & Max & Mean & Std. Dev. & Skewness & Kurtosis & JB$\times10^{-6}$ \\
\cmidrule(lr){1-1}\cmidrule(lr){2-7}\cmidrule(lr){8-13}
Electronic Technology & 80542954 & 430265.82 & 1364898.01 & 38.23 & 1889.12 & 1298.64 & 2355 & 11.65 & 94.39 & 10.52 & 129.45 & 6.25  \\
Non-Energy Minerals & 4772977 & 47789.1 & 120300.14 & 17.87 & 496.12 & 89.88 & 249 & 1.06 & 9.68 & 12.68 & 198.61 & 14.56  \\
Health Technology & 31536002 & 173898.02 & 672731.54 & 32.78 & 1286.22 & 602.58 & 1343 & 6.72 & 52.97 & 10.82 & 140.37 & 7.33  \\
Distribution services & 1553772 & 14435.57 & 33421.05 & 21.97 & 754.79 & 207.67 & 269 & 0.59 & 6.14 & 21.05 & 659.37 & 158.59  \\
Finance & 45639830 & 442359.14 & 1145068.11 & 22.48 & 723.57 & 190.94 & 3476 & 14.66 & 114.54 & 11.66 & 182.00 & 12.23  \\
Technology Services & 58768687 & 198602.17 & 908170.62 & 47.17 & 2643.35 & 2541.67 & 2138 & 8.09 & 68.63 & 12.00 & 193.73 & 13.84  \\
Process Industries & 3574375 & 51038.8 & 113420.64 & 17.3 & 427.37 & 66.79 & 591 & 2.20 & 18.01 & 12.66 & 224.37 & 18.52  \\
Commercial services & 369917 & 18613.36 & 24117.66 & 5.76 & 44.28 & 0.76 & 215 & 0.92 & 8.15 & 12.26 & 181.72 & 12.21  \\
Utilities & 7148018 & 76344.43 & 174961.72 & 20.75 & 637.42 & 148.23 & 836 & 2.73 & 23.20 & 14.98 & 322.98 & 38.22  \\
Health Services & 4330320 & 32169.87 & 87986.03 & 31.45 & 1316.22 & 630.82 & 345 & 1.68 & 13.90 & 11.70 & 163.72 & 9.94  \\
Retail Trade & 15952647 & 147346.78 & 340712.62 & 25.36 & 926.35 & 312.68 & 942 & 5.67 & 47.34 & 10.88 & 135.44 & 6.84  \\
Energy Minerals & 12117418 & 115305.52 & 263236.58 & 24.94 & 920.34 & 308.62 & 636 & 3.90 & 31.65 & 10.69 & 127.30 & 6.05  \\
Consumer services & 26011209 & 149917.34 & 493367.72 & 38.38 & 1780.23 & 1153.49 & 1408 & 6.91 & 54.71 & 10.79 & 140.27 & 7.32  \\
Consumer non-durables & 17447157 & 137296.42 & 381029.69 & 22.94 & 765.24 & 213.5 & 711 & 3.63 & 29.27 & 10.92 & 141.44 & 7.44  \\
Producer Manufacturing & 22102513 & 117743.99 & 424997.45 & 31.07 & 1275.47 & 592.41 & 1164 & 3.77 & 32.68 & 13.40 & 267.79 & 26.31  \\
Industrial Services & 7168024 & 66441.23 & 142155.51 & 24.67 & 951.03 & 329.46 & 406 & 1.60 & 13.22 & 13.32 & 234.39 & 20.22  \\
Communications & 6464633 & 57034.49 & 164533 & 20.41 & 616.6 & 138.73 & 188 & 0.96 & 8.19 & 11.79 & 162.91 & 9.84  \\
Transportation & 2705466 & 27104.64 & 65832.05 & 22.33 & 689.5 & 173.44 & 213 & 0.98 & 8.93 & 13.10 & 208.70 & 16.07  \\
Consumer durables & 8070407 & 81704.94 & 196434.14 & 19.71 & 594.8 & 129.09 & 555 & 2.13 & 18.33 & 12.02 & 191.27 & 13.50  \\
\hline
\end{tabular}}
\caption{\footnotesize{Summary statistics of the exogenous regressors belonging to the Information Volumes $\mathcal{M}_{\sI\sV}$ regressor set.}}
 \label{tab:IV_summary_stat}
%\end{table}
\end{sidewaystable}
%%
%
%
%::::::::::::::::::::::::::::::::::::::::::::::::::::::::::::::::::::::
% TABLE: DATA SUMMARY STATISTICS of
% EXOGENOUS REGRESSORS SE
%::::::::::::::::::::::::::::::::::::::::::::::::::::::::::::::::::::::
%%
\begin{sidewaystable}[!ht]
\captionsetup{font={small}, labelfont=sc}
%\begin{table}[ht]
\centering
 \smallskip
  \centering
   \resizebox{1.0\columnwidth}{!}{%
    \setlength\tabcolsep{1mm}
 \begin{tabular}{lccccccccccccccccccc}\\
  \hline
 & \multicolumn{6}{c}{HIGH} & \multicolumn{6}{c}{POS}  & \multicolumn{6}{c}{NEG}\\
\cmidrule(lr){1-1}\cmidrule(lr){2-7}\cmidrule(lr){8-13}\cmidrule(lr){14-19}
   Name & Max & Mean & Std. Dev. & Skewness & Kurtosis & JB$\times10^{-6}$ & Max & Mean & Std. Dev. & Skewness & Kurtosis & JB$\times10^{-6}$ & Max & Mean & Std. Dev. & Skewness & Kurtosis & JB$\times10^{-6}$ \\
\cmidrule(lr){1-1}\cmidrule(lr){2-7}\cmidrule(lr){8-13}\cmidrule(lr){14-19}
Electronic Technology & 178 & 0.96 & 7.93 & 11.1 & 144.6 & 7.78 & 87 & 0.49 & 4.22 & 10.87 & 130.93 & 6.4 & 88 & 0.44 & 3.72 & 10.89 & 136.66 & 6.96  \\
Non-Energy Minerals & 31 & 0.13 & 1.27 & 13.23 & 204.89 & 15.51 & 18 & 0.07 & 0.65 & 13.38 & 221.72 & 18.12 & 18 & 0.06 & 0.62 & 15.92 & 303.92 & 33.92  \\
Health Technology & 105 & 0.53 & 4.57 & 12.28 & 178.44 & 11.79 & 69 & 0.32 & 2.84 & 12.18 & 173.96 & 11.21 & 32 & 0.19 & 1.69 & 11.04 & 135.17 & 6.82  \\
Distribution services & 40 & 0.06 & 0.81 & 24.82 & 879.51 & 281.92 & 8 & 0.02 & 0.29 & 15.79 & 295.05 & 31.99 & 16 & 0.02 & 0.28 & 29.91 & 1321.7 & 635.93  \\
Finance & 320 & 1.44 & 12.12 & 11.51 & 160.96 & 9.6 & 140 & 0.69 & 5.75 & 11.58 & 161.62 & 9.68 & 177 & 0.71 & 6.06 & 12.25 & 195.93 & 14.16  \\
Technology Services & 146 & 0.53 & 4.46 & 12.83 & 231.45 & 19.7 & 66 & 0.3 & 2.62 & 11.75 & 164.91 & 10.08 & 90 & 0.3 & 2.59 & 13.46 & 261.15 & 25.04  \\
Process Industries & 53 & 0.22 & 2.08 & 14.36 & 256.42 & 24.19 & 21 & 0.09 & 0.87 & 13.59 & 223.37 & 18.4 & 19 & 0.09 & 0.85 & 12.85 & 192.73 & 13.73  \\
Commercial services & 31 & 0.07 & 0.75 & 18.43 & 491.65 & 88.31 & 13 & 0.04 & 0.42 & 15.28 & 290.35 & 30.97 & 12 & 0.03 & 0.40 & 16.44 & 335.07 & 41.18  \\
Utilities & 71 & 0.19 & 1.96 & 18.67 & 461.07 & 77.74 & 69 & 0.09 & 1.07 & 36.35 & 2046.68 & 1523.71 & 25 & 0.08 & 0.75 & 16.13 & 339.99 & 42.37  \\
Health Services & 38 & 0.12 & 1.18 & 15.25 & 304.95 & 34.12 & 22 & 0.07 & 0.69 & 15.06 & 288.78 & 30.63 & 13 & 0.06 & 0.56 & 13.35 & 207.19 & 15.85  \\
Retail Trade & 132 & 0.61 & 5.59 & 12.68 & 194.57 & 13.99 & 63 & 0.33 & 3.11 & 12 & 164.65 & 10.06 & 47 & 0.24 & 2.24 & 12.31 & 177.21 & 11.63  \\
Energy Minerals & 90 & 0.42 & 4 & 12.85 & 186.96 & 12.94 & 32 & 0.15 & 1.39 & 13.53 & 218.3 & 17.58 & 40 & 0.19 & 1.79 & 13.10 & 203.21 & 15.25  \\
Consumer services & 99 & 0.46 & 3.86 & 11.54 & 160.65 & 9.57 & 66 & 0.26 & 2.24 & 11.87 & 181.43 & 12.16 & 55 & 0.23 & 1.94 & 11.73 & 174.29 & 11.24  \\
Consumer non-durables & 78 & 0.31 & 2.86 & 13.79 & 237.36 & 20.74 & 35 & 0.15 & 1.31 & 12.14 & 180.09 & 12 & 36 & 0.12 & 1.15 & 14.34 & 261.66 & 25.17  \\
Producer Manufacturing & 99 & 0.33 & 3.15 & 14.72 & 281.97 & 29.2 & 47 & 0.14 & 1.31 & 14.54 & 302.54 & 33.56 & 39 & 0.16 & 1.49 & 13.11 & 208.06 & 15.98  \\
Industrial Services & 50 & 0.16 & 1.59 & 16.6 & 356.19 & 46.49 & 32 & 0.07 & 0.74 & 21.03 & 623.38 & 141.82 & 18 & 0.07 & 0.69 & 15.17 & 282.6 & 29.35  \\
Communications & 53 & 0.11 & 1.22 & 21.1 & 643.89 & 151.27 & 44 & 0.09 & 0.91 & 23.24 & 840.77 & 257.59 & 11 & 0.03 & 0.36 & 16.21 & 330.23 & 40.00  \\
Transportation & 29 & 0.09 & 0.95 & 16.26 & 332.13 & 40.46 & 13 & 0.04 & 0.46 & 16.63 & 336.18 & 41.46 & 23 & 0.05 & 0.54 & 20.90 & 615.66 & 138.34  \\
Consumer durables & 69 & 0.22 & 2.11 & 14.25 & 270.56 & 26.89 & 24 & 0.09 & 0.87 & 13.46 & 231.23 & 19.69 & 24 & 0.08 & 0.78 & 14.30 & 258.08 & 24.49  \\
\hline
\end{tabular}}
\caption{\footnotesize{Summary statistics of the exogenous regressors belonging to the Sentiment $\mathcal{M}_{\sS\sE}$ regressor set.}}
\label{tab:SE_summary_stat}
%\end{table}
\end{sidewaystable}
%
%
%::::::::::::::::::::::::::::::::::::::::::::::::::::::::::::::::::::::
% TABLE: A/E
%::::::::::::::::::::::::::::::::::::::::::::::::::::::::::::::::::::::
\begin{sidewaystable}[!ht]
\captionsetup{font={small}, labelfont=sc}
%\begin{table}[ht]
\centering
 \smallskip
   \resizebox{0.90\columnwidth}{!}{%
    \setlength\tabcolsep{1mm}
\begin{tabular}{lcccccccccccccccccccccccccccc}
    \toprule
 \multirow{3}{*}{$\tau=1\%$}& \multicolumn{9}{c}{GED} & \multicolumn{9}{c}{Gaussian} & \multicolumn{9}{c}{Student--t}\\
\cmidrule(lr){2-10}\cmidrule(lr){11-19}\cmidrule(lr){20-28}
 & \multicolumn{3}{c}{EGARCH} & \multicolumn{3}{c}{GJR-GARCH} & \multicolumn{3}{c}{GARCH} & \multicolumn{3}{c}{EGARCH} & \multicolumn{3}{c}{GJR-GARCH} & \multicolumn{3}{c}{GARCH} & \multicolumn{3}{c}{EGARCH} & \multicolumn{3}{c}{GJR-GARCH} & \multicolumn{3}{c}{GARCH} \\
 \cmidrule(lr){2-4}\cmidrule(lr){5-7}\cmidrule(lr){8-10}\cmidrule(lr){11-13}\cmidrule(lr){14-16}\cmidrule(lr){17-19}\cmidrule(lr){20-22}\cmidrule(lr){23-25}\cmidrule(lr){26-28}
  & $\mathcal{M}_{\sS\sE}$ & $\mathcal{M}_{\sI\sV}$ & $\mathcal{M}_{\sN}$ & $\mathcal{M}_{\sS\sE}$ & $\mathcal{M}_{\sI\sV}$ & $\mathcal{M}_{\sN}$ & $\mathcal{M}_{\sS\sE}$ & $\mathcal{M}_{\sI\sV}$ & $\mathcal{M}_{\sN}$ & $\mathcal{M}_{\sS\sE}$ & $\mathcal{M}_{\sI\sV}$ & $\mathcal{M}_{\sN}$ & $\mathcal{M}_{\sS\sE}$ & $\mathcal{M}_{\sI\sV}$ & $\mathcal{M}_{\sN}$ & $\mathcal{M}_{\sS\sE}$ & $\mathcal{M}_{\sI\sV}$ & $\mathcal{M}_{\sN}$ & $\mathcal{M}_{\sS\sE}$ & $\mathcal{M}_{\sI\sV}$ & $\mathcal{M}_{\sN}$ & $\mathcal{M}_{\sS\sE}$ & $\mathcal{M}_{\sI\sV}$ & $\mathcal{M}_{\sN}$ & $\mathcal{M}_{\sS\sE}$ & $\mathcal{M}_{\sI\sV}$ & $\mathcal{M}_{\sN}$ \\
\hline
Commercial services &  0.60  &  0.72  &  0.58  &  0.65  &  0.81  &  0.60  &  0.60  &  0.93  &  0.58  &  0.77  &  0.77  &  0.81  &  0.84  &  0.79  &  0.86  &  0.77  & {\colorred\textbf{ 0.98 }} &  0.84  &  0.67  &  0.74  &  0.67  &  0.70  &  0.58  &  0.77  &  0.88  &  0.58  &  0.77  \\
Communications &  0.60  &  0.77  &  0.65  &  2.67  &  0.70  &  2.84  &  4.40  &  1.14  &  4.63  &  0.88  &  0.91  &  0.93  &  0.91  &  0.72  &  0.98  & {\colorred\textbf{ 1.00 }} &  0.88  &  1.16  &  0.67  &  0.84  &  0.74  &  0.79  &  0.53  &  0.74  &  0.81  &  0.51  &  0.77  \\
Consumer durables &  0.67  &  0.58  &  0.67  &  2.42  &  1.16  &  2.65  &  4.21  &  0.74  &  4.65  &  1.16  &  1.12  &  1.23  & {\colorred\textbf{ 1.02 }} &  0.60  &  1.19  &  1.37  &  0.72  &  1.37  &  0.79  &  0.74  &  0.79  &  0.77  &  0.58  &  0.72  &  0.74  &  0.63  &  0.72  \\
Consumer non-durables &  0.58  &  0.70  &  0.56  &  0.67  &  0.74  &  0.77  &  0.74  &  0.86  &  0.93  &  0.81  &  1.23  &  0.86  &  0.95  &  0.86  &  1.14  & {\colorred\textbf{ 0.98 }} & {\colorred\textbf{ 0.98 }} &  1.12  &  0.56  &  0.70  &  0.53  &  0.84  &  0.58  & {\colorred\textbf{ 0.98 }} &  0.93  &  0.67  &  1.09  \\
Consumer services &  0.72  &  0.95  &  0.74  &  0.67  &  0.77  &  0.70  &  2.77  &  0.95  &  1.37  &  1.07  &  1.40  &  1.05  &  0.98  &  0.67  & {\colorred\textbf{ 1.00 }} &  1.02  &  0.84  & {\colorred\textbf{ 1.00 }} &  0.88  & {\colorred\textbf{ 1.00 }} &  0.81  &  0.86  &  0.72  &  0.91  &  0.84  &  0.77  &  0.86  \\
Distribution services &  0.58  &  0.67  &  0.63  &  0.86  &  0.67  &  0.83  &  2.95  &  0.81  &  1.14  &  0.72  & {\colorred\textbf{ 0.91 }} &  0.79  &  1.12  &  0.63  &  1.37  &  1.16  &  0.84  &  1.49  &  0.65  &  0.65  &  0.70  &  0.79  &  0.65  &  0.72  & {\colorred\textbf{ 0.91 }} &  0.65  &  0.84  \\
Electronic Technology &  0.81  & 0.00 &  0.79  &  0.81  &  0.70  &  0.78  &  0.81  &  0.84  &  0.79  & {\colorred\textbf{ 0.95 }} &  0.51  &  0.93  &  0.93  &  0.67  &  1.09  &  0.93  &  0.72  &  1.09  &  0.84  &  0.47  &  0.79  & {\colorred\textbf{ 0.95 }} &  0.70  &  0.86  &  0.88  &  0.65  &  0.84  \\
Energy Minerals &  0.42  &  0.67  &  0.65  &  0.58  &  0.74  &  0.60  &  0.58  &  0.77  &  0.60  &  0.72  & {\colorred\textbf{ 0.98 }} &  0.91  &  0.77  &  0.77  &  1.14  &  0.79  &  0.88  &  1.19  &  0.56  &  0.56  &  0.70  &  0.65  &  0.44  &  0.72  &  0.65  &  0.44  &  0.74  \\
Finance &  0.58  &  0.26  &  0.60  &  0.74  &  0.47  &  0.58  &  0.77  &  0.49  &  0.60  &  0.88  &  0.63  &  0.93  &  0.98  &  0.49  &  0.93  &  0.93  &  0.56  &  0.77  &  0.60  &  0.21  &  0.67  &  0.88  &  0.40  &  0.88  & {\colorred\textbf{ 1.00 }} &  0.47  &  0.88  \\
Health Services &  0.44  &  0.53  &  0.47  &  0.44  &  0.49  &  0.49  &  0.51  &  0.53  &  0.49  &  0.65  &  0.70  &  0.63  &  0.70  &  0.49  &  0.72  &  0.56  &  0.51  &  0.60  &  0.47  &  0.51  &  0.49  &  0.58  &  0.37  &  0.56  & {\colorred\textbf{ 0.77 }} &  0.37  &  0.53  \\
Health Technology &  0.63  &  0.26  &  0.63  &  0.72  &  0.49  &  0.79  &  2.65  &  0.58  &  0.88  &  0.81  &  0.91  &  0.81  &  0.88  &  0.51  &  0.88  & {\colorred\textbf{ 0.98 }} &  0.65  &  0.84  &  0.67  &  0.56  &  0.67  &  0.77  &  0.53  &  0.93  &  0.88  &  0.47  &  1.07  \\
Industrial Services &  0.47  &  0.40  &  0.44  &  0.47  &  0.65  &  0.47  &  0.47  &  0.74  &  0.49  &  0.63  &  0.84  &  0.67  &  0.70  &  0.72  &  0.86  &  0.79  &  0.74  & {\colorred\textbf{ 0.95 }} &  0.47  &  0.42  &  0.56  &  0.53  &  0.42  &  0.56  &  0.51  &  0.42  &  0.56  \\
Non-Energy Minerals &  0.37  &  0.49  &  0.44  &  0.42  &  0.51  &  0.37  &  2.58  &  0.67  & {\colorred\textbf{ 0.91 }} &  0.44  &  0.72  &  0.49  &  0.67  &  0.44  &  0.67  &  0.84  &  0.63  &  0.84  &  0.44  &  0.49  &  0.49  &  0.51  &  0.35  &  0.53  &  0.49  &  0.28  &  0.51  \\
Process Industries &  0.58  &  0.65  &  0.65  &  0.60  &  0.77  &  0.60  &  0.65  &  0.84  &  0.63  & {\colorred\textbf{ 0.98 }} &  1.23  & {\colorred\textbf{ 0.98 }} &  1.07  &  0.72  &  1.07  &  1.02  &  0.84  &  1.07  &  0.65  &  0.63  &  0.58  &  0.60  &  0.51  &  0.67  &  0.67  &  0.67  &  0.70  \\
Producer Manufacturing &  0.53  &  0.37  &  0.49  &  0.51  &  0.58  &  0.51  &  0.56  &  0.63  &  0.49  &  0.67  &  0.70  &  0.67  &  0.67  &  0.63  & {\colorred\textbf{ 0.72 }} & {\colorred\textbf{ 0.72 }} &  0.67  &  0.65  &  0.56  &  0.42  &  0.56  &  0.67  &  0.51  &  0.58  &  0.67  &  0.53  &  0.58  \\
Retail Trade &  0.86  &  1.26  &  0.93  &  0.84  &  0.79  & {\colorred\textbf{ 0.98 }} &  0.86  &  0.84  &  1.16  &  1.44  &  1.35  &  1.44  &  1.30  &  0.79  &  1.30  &  1.21  & {\colorred\textbf{ 0.98 }} &  1.37  & {\colorred\textbf{ 0.98 }} &  1.09  &  0.93  &  1.02  &  0.70  &  1.14  &  1.14  &  0.74  &  1.23  \\
Technology Services &  0.70  & 0.00 &  0.72  &  0.74  &  0.65  &  0.70  &  0.72  &  0.86  &  0.77  &  0.88  &  0.77  &  0.88  &  0.86  &  0.63  & {\colorred\textbf{ 1.05 }} &  0.91  &  0.72  &  0.93  &  0.84  &  0.56  &  0.79  &  0.86  &  0.60  &  0.81  &  0.84  &  0.67  &  0.91  \\
Transportation &  0.77  &  1.02  &  0.79  &  0.88  &  1.14  &  0.79  &  3.56  &  1.35  &  1.49  &  1.12  &  1.51  &  1.14  &  1.30  &  1.09  &  1.09  &  1.21  &  1.19  &  1.21  &  0.81  & {\colorred\textbf{ 1.00 }} &  0.86  & {\colorred\textbf{ 1.00 }} &  0.84  &  0.93  &  0.98  &  0.74  &  1.07  \\
Utilities &  1.09  &  1.12  & {\colorred\textbf{ 1.00 }} &  1.16  &  1.26  &  1.23  &  1.19  &  1.60  &  1.23  &  1.58  &  1.81  &  1.53  &  1.42  &  1.51  &  1.58  &  1.79  &  1.91  &  1.47  &  0.91  &  1.26  &  0.86  &  1.28  &  0.72  &  1.44  &  1.30  &  0.95  &  1.51  \\
%%%%
 \toprule
\multirow{3}{*}{$\tau=0.1\%$}& \multicolumn{9}{c}{GED} & \multicolumn{9}{c}{Gaussian} & \multicolumn{9}{c}{Student--t}\\
\cmidrule(lr){2-10}\cmidrule(lr){11-19}\cmidrule(lr){20-28}
 & \multicolumn{3}{c}{EGARCH} & \multicolumn{3}{c}{GJR-GARCH} & \multicolumn{3}{c}{GARCH} & \multicolumn{3}{c}{EGARCH} & \multicolumn{3}{c}{GJR-GARCH} & \multicolumn{3}{c}{GARCH} & \multicolumn{3}{c}{EGARCH} & \multicolumn{3}{c}{GJR-GARCH} & \multicolumn{3}{c}{GARCH} \\
 \cmidrule(lr){2-4}\cmidrule(lr){5-7}\cmidrule(lr){8-10}\cmidrule(lr){11-13}\cmidrule(lr){14-16}\cmidrule(lr){17-19}\cmidrule(lr){20-22}\cmidrule(lr){23-25}\cmidrule(lr){26-28}
  & $\mathcal{M}_{\sS\sE}$ & $\mathcal{M}_{\sI\sV}$ & $\mathcal{M}_{\sN}$ & $\mathcal{M}_{\sS\sE}$ & $\mathcal{M}_{\sI\sV}$ & $\mathcal{M}_{\sN}$ & $\mathcal{M}_{\sS\sE}$ & $\mathcal{M}_{\sI\sV}$ & $\mathcal{M}_{\sN}$ & $\mathcal{M}_{\sS\sE}$ & $\mathcal{M}_{\sI\sV}$ & $\mathcal{M}_{\sN}$ & $\mathcal{M}_{\sS\sE}$ & $\mathcal{M}_{\sI\sV}$ & $\mathcal{M}_{\sN}$ & $\mathcal{M}_{\sS\sE}$ & $\mathcal{M}_{\sI\sV}$ & $\mathcal{M}_{\sN}$ & $\mathcal{M}_{\sS\sE}$ & $\mathcal{M}_{\sI\sV}$ & $\mathcal{M}_{\sN}$ & $\mathcal{M}_{\sS\sE}$ & $\mathcal{M}_{\sI\sV}$ & $\mathcal{M}_{\sN}$ & $\mathcal{M}_{\sS\sE}$ & $\mathcal{M}_{\sI\sV}$ & $\mathcal{M}_{\sN}$ \\
\hline
%%%%
Commercial services &  2.50  &  2.00  &  2.75  &  3.00  &  3.50  &  3.00  &  3.00  &  3.50  &  3.25  &  4.25  &  3.25  &  4.75  &  4.25  &  3.50  &  4.50  &  4.75  &  3.50  &  5.00  &  2.25  & {\colorred\textbf{ 1.25 }} &  2.00  &  2.00  &  1.50  &  2.00  &  3.00  &  1.75  &  2.25  \\
Communications &  2.25  &  2.00  &  2.25  &  13.00  &  3.50  &  14.50  &  22.75  &  5.50  &  23.75  &  3.75  &  3.00  &  4.25  &  4.50  &  3.50  &  4.75  &  4.75  &  3.75  &  5.00  &  1.75  &  1.25  &  1.75  &  2.25  & {\colorred\textbf{ 1.00 }} &  1.75  &  2.25  & {\colorred\textbf{ 1.00 }} &  1.75  \\
Consumer durables &  2.75  &  0.50  &  3.00  &  10.75  &  3.50  &  12.50  &  19.50  &  2.50  &  22.50  &  4.75  &  2.50  &  4.75  &  5.50  &  2.25  &  6.25  &  6.75  &  3.00  &  7.00  &  2.50  & {\colorred\textbf{ 0.75 }} &  2.00  &  2.50  & {\colorred\textbf{ 0.75 }} &  2.25  &  2.50  &  0.25  &  2.00  \\
Consumer non-durables &  2.75  &  2.25  &  2.50  &  3.25  &  3.50  &  2.75  &  2.50  &  3.75  &  3.00  &  4.00  &  3.75  &  4.00  &  5.00  &  4.00  &  5.00  &  4.75  &  4.75  &  4.50  & {\colorred\textbf{ 1.25 }} &  1.50  &  1.50  &  2.00  & {\colorred\textbf{ 0.75 }} &  2.75  &  2.25  & {\colorred\textbf{ 1.25 }} &  2.50  \\
Consumer services &  3.00  &  1.50  &  3.25  &  3.00  &  2.75  &  3.00  &  24.00  &  3.00  &  4.75  &  4.25  &  5.25  &  4.75  &  5.25  &  3.25  &  5.25  &  5.00  &  3.75  &  5.50  &  2.25  & {\colorred\textbf{ 1.00 }} &  2.00  &  2.50  &  1.75  &  2.25  &  2.25  &  1.75  &  2.50  \\
Distribution services &  2.50  & {\colorred\textbf{ 1.00 }} &  2.25  &  5.75  &  3.25  &  4.20  &  24.50  &  3.25  &  5.00  &  4.25  &  2.25  &  4.75  &  7.00  &  3.75  &  7.00  &  7.25  &  4.00  &  7.50  &  2.00  &  0.75  &  2.00  &  1.75  &  0.50  &  2.25  &  2.00  &  0.75  &  2.75  \\
Electronic Technology &  3.25  & 0.00 &  3.00  &  3.50  &  3.25  &  3.25  &  3.50  &  3.50  &  3.25  &  5.00  & {\colorred\textbf{ 1.00 }} &  5.75  &  5.75  &  3.25  &  5.50  &  5.75  &  3.50  &  5.50  &  2.00  &  0.50  &  1.50  &  2.00  &  1.75  &  2.00  &  2.25  &  1.25  &  2.00  \\
Energy Minerals &  2.00  & {\colorred\textbf{ 1.00 }} &  2.25  &  2.00  &  2.50  &  2.50  &  2.00  &  2.25  &  2.50  &  3.50  &  3.50  &  4.25  &  3.75  &  2.75  &  4.75  &  4.00  &  3.00  &  4.75  &  1.50  &  0.50  & {\colorred\textbf{ 1.00 }} &  1.75  &  0.50  & {\colorred\textbf{ 1.00 }} &  1.75  &  0.25  & {\colorred\textbf{ 1.00 }} \\
Finance &  3.25  &  0.50  &  3.25  &  3.25  &  3.00  &  3.50  &  3.50  &  3.00  &  3.25  &  4.00  &  1.75  &  4.25  &  6.00  &  3.25  &  4.50  &  6.00  &  3.75  &  4.50  &  2.00  & {\colorred\textbf{ 0.75 }} &  1.75  &  2.50  &  1.75  &  2.25  &  3.00  & {\colorred\textbf{ 1.25 }} &  2.25  \\
Health Services &  2.50  &  0.75  &  2.00  &  2.25  &  3.00  &  2.00  &  2.00  &  3.00  &  2.00  &  3.25  &  2.25  &  3.00  &  3.25  &  3.00  &  3.25  &  3.50  &  3.25  &  3.25  &  1.25  &  0.75  & {\colorred\textbf{ 1.00 }} &  1.25  &  0.50  &  1.25  &  2.75  &  0.50  & {\colorred\textbf{ 1.00 }} \\
Health Technology &  2.50  &  0.75  &  2.75  &  2.25  &  2.25  &  2.50  &  23.25  &  2.00  &  2.75  &  4.00  &  3.75  &  4.00  &  4.00  &  2.25  &  4.75  &  4.25  &  2.25  &  4.75  &  1.25  &  1.25  & {\colorred\textbf{ 1.00 }} &  1.75  &  0.50  &  2.50  &  2.75  & {\colorred\textbf{ 1.00 }} &  2.50  \\
Industrial Services &  2.00  & {\colorred\textbf{ 1.00 }} &  1.75  &  2.00  &  2.25  &  1.75  &  2.00  &  2.75  &  1.75  &  3.25  &  1.50  &  3.50  &  3.75  &  2.50  &  3.50  &  4.25  &  2.25  &  3.50  &  1.50  & {\colorred\textbf{ 1.00 }} & {\colorred\textbf{ 1.00 }} &  1.50  &  0.50  & {\colorred\textbf{ 1.00 }} &  1.75  &  0.50  & {\colorred\textbf{ 1.00 }} \\
Non-Energy Minerals &  1.75  & {\colorred\textbf{ 1.00 }} &  1.75  &  2.00  &  2.00  &  2.00  &  22.00  &  2.50  &  3.50  &  3.00  &  1.75  &  3.00  &  4.00  &  2.25  &  3.75  &  3.75  &  2.50  &  3.50  &  1.75  &  0.75  &  1.50  &  1.75  & {\colorred\textbf{ 1.00 }} &  1.50  &  1.75  & {\colorred\textbf{ 1.00 }} &  1.50  \\
Process Industries &  2.75  & {\colorred\textbf{ 1.25 }} &  2.75  &  3.00  &  4.00  &  3.00  &  3.00  &  3.25  &  2.75  &  4.50  &  3.50  &  4.00  &  4.75  &  3.75  &  4.50  &  4.50  &  3.50  &  4.25  &  1.75  & {\colorred\textbf{ 0.75 }} &  1.75  &  2.00  &  1.50  &  2.00  &  2.25  & {\colorred\textbf{ 0.75 }} &  2.00  \\
Producer Manufacturing &  3.00  &  1.50  &  3.00  &  3.25  &  2.50  &  2.75  &  2.75  &  2.75  &  2.75  &  4.50  &  2.00  &  4.50  &  4.50  &  3.25  &  4.75  &  4.25  &  3.00  &  3.75  &  1.75  & {\colorred\textbf{ 1.00 }} &  1.50  &  2.50  & {\colorred\textbf{ 1.00 }} &  2.00  &  2.00  & {\colorred\textbf{ 1.00 }} &  2.00  \\
Retail Trade &  3.75  &  1.75  &  3.50  &  3.50  &  3.25  &  3.75  &  4.00  &  3.25  &  4.25  &  5.75  &  4.25  &  6.25  &  6.25  &  3.75  &  6.25  &  7.00  &  4.75  &  7.00  &  3.00  & {\colorred\textbf{ 1.00 }} &  3.00  &  2.75  &  1.25  &  3.50  &  3.25  &  1.25  &  3.50  \\
Technology Services &  3.00  & 0.00 &  2.75  &  3.00  &  3.50  &  2.75  &  2.75  &  4.50  &  2.75  &  4.75  &  3.00  &  4.75  &  4.50  &  4.50  &  4.75  &  5.25  &  4.25  &  4.50  &  2.50  &  0.50  &  2.00  &  2.50  &  1.25  &  2.00  &  2.50  & {\colorred\textbf{ 1.00 }} &  2.50  \\
Transportation &  2.25  &  2.25  &  2.50  &  2.00  &  3.25  &  2.00  &  31.25  &  4.75  &  5.25  &  5.00  &  4.50  &  6.00  &  5.25  &  3.50  &  5.50  &  5.00  &  4.75  &  7.00  &  2.00  & {\colorred\textbf{ 1.00 }} &  1.75  &  1.75  & {\colorred\textbf{ 1.00 }} &  2.00  &  1.75  & {\colorred\textbf{ 1.00 }} &  1.75  \\
Utilities &  3.75  &  2.00  &  3.50  &  4.00  &  3.50  &  3.75  &  4.25  &  3.75  &  4.00  &  5.50  &  5.75  &  6.75  &  6.00  &  4.50  &  6.25  &  8.50  &  5.50  &  7.25  &  1.75  &  1.50  &  1.25  &  3.00  & {\colorred\textbf{ 1.00 }} &  3.25  &  2.75  & {\colorred\textbf{ 1.00 }} &  3.25  \\
\bottomrule
\end{tabular}}
\caption{\footnotesize{Actual over expected (A/E) VaR exceedance rates for all models at confidence levels $\tau=\left(1\%,0.1\%\right)$. For each sector, in {\colorred\textbf{bold}} we denote the A/E value closest to one.}}
\label{tab:A/E_all}
%
%\end{table}
\end{sidewaystable}

%::::::::::::::::::::::::::::::::::::::::::::::::::::::::::::::::::::::
% TABLE: AD
%::::::::::::::::::::::::::::::::::::::::::::::::::::::::::::::::::::::
\begin{sidewaystable}[htbp]
\captionsetup{font={small}, labelfont=sc}
 \smallskip
  \centering
   \resizebox{0.90\columnwidth}{!}{%
    \setlength\tabcolsep{1mm}
\begin{tabular}{lcccccccccccccccccccccccccccc}
    \toprule
 \multirow{3}{*}{$\tau=1\%$}& \multicolumn{9}{c}{GED} & \multicolumn{9}{c}{Gaussian} & \multicolumn{9}{c}{Student--t}\\
\cmidrule(lr){2-10}\cmidrule(lr){11-19}\cmidrule(lr){20-28}
 & \multicolumn{3}{c}{EGARCH} & \multicolumn{3}{c}{GJR-GARCH} & \multicolumn{3}{c}{GARCH} & \multicolumn{3}{c}{EGARCH} & \multicolumn{3}{c}{GJR-GARCH} & \multicolumn{3}{c}{GARCH} & \multicolumn{3}{c}{EGARCH} & \multicolumn{3}{c}{GJR-GARCH} & \multicolumn{3}{c}{GARCH} \\
 \cmidrule(lr){2-4}\cmidrule(lr){5-7}\cmidrule(lr){8-10}\cmidrule(lr){11-13}\cmidrule(lr){14-16}\cmidrule(lr){17-19}\cmidrule(lr){20-22}\cmidrule(lr){23-25}\cmidrule(lr){26-28}
  & $\mathcal{M}_{\sS\sE}$ & $\mathcal{M}_{\sI\sV}$ & $\mathcal{M}_{\sN}$ & $\mathcal{M}_{\sS\sE}$ & $\mathcal{M}_{\sI\sV}$ & $\mathcal{M}_{\sN}$ & $\mathcal{M}_{\sS\sE}$ & $\mathcal{M}_{\sI\sV}$ & $\mathcal{M}_{\sN}$ & $\mathcal{M}_{\sS\sE}$ & $\mathcal{M}_{\sI\sV}$ & $\mathcal{M}_{\sN}$ & $\mathcal{M}_{\sS\sE}$ & $\mathcal{M}_{\sI\sV}$ & $\mathcal{M}_{\sN}$ & $\mathcal{M}_{\sS\sE}$ & $\mathcal{M}_{\sI\sV}$ & $\mathcal{M}_{\sN}$ & $\mathcal{M}_{\sS\sE}$ & $\mathcal{M}_{\sI\sV}$ & $\mathcal{M}_{\sN}$ & $\mathcal{M}_{\sS\sE}$ & $\mathcal{M}_{\sI\sV}$ & $\mathcal{M}_{\sN}$ & $\mathcal{M}_{\sS\sE}$ & $\mathcal{M}_{\sI\sV}$ & $\mathcal{M}_{\sN}$ \\
\hline
Commercial services &  0.73  &  0.53  &  0.76  &  0.71  &  0.72  &  0.73  &  0.72  &  0.69  &  0.82  &  0.49  & {\colorred\textbf{ 0.42 }} &  0.75  &  0.72  &  0.60  &  0.71  &  0.60  &  0.53  &  0.75  &  0.74  &  0.56  &  0.78  &  0.73  &  0.50  &  0.78  &  0.82  &  0.51  &  0.76  \\
Communications &  1.04  & {\colorred\textbf{ 0.76 }} &  1.03  &  1.17  &  1.04  &  1.17  &  1.20  &  1.02  &  1.20  &  1.06  &  0.77  &  1.02  &  1.06  &  1.02  &  1.07  &  1.08  &  1.01  &  1.11  &  1.06  &  0.78  &  1.06  &  1.09  &  0.98  &  1.07  &  1.09  &  0.99  &  1.08  \\
Consumer durables &  0.64  & {\colorred\textbf{ 0.28 }} &  0.63  &  0.70  &  0.62  &  0.70  &  0.72  &  0.49  &  0.78  &  0.64  &  0.30  &  0.64  &  0.66  &  0.44  &  0.68  &  0.66  &  0.50  &  0.70  &  0.65  &  0.28  &  0.64  &  0.65  &  0.43  &  0.64  &  0.63  &  0.42  &  0.63  \\
Consumer non-durables &  0.32  &  0.32  &  0.32  &  0.28  &  0.29  &  0.30  &  0.30  &  0.28  &  0.32  &  0.33  &  0.31  &  0.33  &  0.29  &  0.28  &  0.32  &  0.31  &  0.28  &  0.33  &  0.32  &  0.28  &  0.32  &  0.30  &  0.31  &  0.31  &  0.32  & {\colorred\textbf{ 0.26 }} &  0.32  \\
Consumer services &  0.88  &  0.41  &  0.87  &  0.88  &  0.85  &  0.87  &  0.90  &  0.80  &  0.89  &  0.90  & {\colorred\textbf{ 0.35 }} &  0.89  &  0.91  &  0.78  &  0.91  &  0.88  &  0.78  &  0.89  &  0.89  &  0.49  &  0.87  &  0.91  &  0.81  &  0.89  &  0.91  &  0.79  &  0.90  \\
Distribution services &  1.41  &  1.19  &  1.40  &  1.42  &  1.39  &  1.34  &  1.43  &  1.34  &  1.42  &  1.42  & {\colorred\textbf{ 1.10 }} &  1.42  &  1.43  &  1.35  &  1.43  &  1.42  &  1.40  &  1.42  &  1.41  &  1.26  &  1.41  &  1.43  &  1.33  &  1.43  &  1.45  &  1.30  &  1.43  \\
Electronic Technology &  1.16  &  1.14  &  1.14  &  1.16  &  0.69  &  0.72  &  1.16  &  0.72  &  1.13  &  1.18  & {\colorred\textbf{ 0.12 }} &  1.18  &  1.18  &  0.68  &  1.17  &  1.19  &  0.68  &  1.17  &  1.17  &  0.43  &  1.15  &  1.18  &  0.72  &  1.15  &  1.17  &  0.67  &  1.15  \\
Energy Minerals &  0.76  & {\colorred\textbf{ 0.53 }} &  0.90  &  0.80  &  0.73  &  0.89  &  0.80  &  0.71  &  0.89  &  0.88  &  0.55  &  0.91  &  0.82  &  0.72  &  0.91  &  0.82  &  0.71  &  0.91  &  0.88  &  0.58  &  0.91  &  0.81  &  0.68  &  0.91  &  0.82  &  0.65  &  0.91  \\
Finance &  0.59  & {\colorred\textbf{ 0.39 }} &  0.59  &  0.63  &  0.59  &  0.61  &  0.64  &  0.59  &  0.61  &  0.61  &  0.59  &  0.61  &  0.64  &  0.59  &  0.62  &  0.65  &  0.58  &  0.61  &  0.60  &  0.42  &  0.60  &  0.65  &  0.60  &  0.62  &  0.66  &  0.58  &  0.62  \\
Health Services &  0.55  & {\colorred\textbf{ 0.27 }} &  0.74  &  0.64  &  0.58  &  0.77  &  0.61  &  0.57  &  0.76  &  0.52  &  0.28  &  0.75  &  0.55  &  0.59  &  0.76  &  0.64  &  0.59  &  0.79  &  0.62  &  0.36  &  0.76  &  0.65  &  0.55  &  0.79  &  0.73  &  0.53  &  0.78  \\
Health Technology &  0.31  &  0.28  &  0.30  &  0.26  &  0.28  &  0.28  &  0.34  &  0.29  &  0.31  &  0.34  &  0.44  &  0.34  &  0.32  &  0.29  &  0.31  &  0.27  &  0.30  &  0.31  &  0.31  &  0.29  &  0.30  & {\colorred\textbf{ 0.26 }} &  0.29  &  0.29  &  0.28  &  0.27  &  0.28  \\
Industrial Services &  0.59  & {\colorred\textbf{ 0.28 }} &  0.63  &  0.45  &  0.38  &  0.63  &  0.52  &  0.40  &  0.62  &  0.58  &  0.31  &  0.64  &  0.42  &  0.46  &  0.65  &  0.48  &  0.38  &  0.67  &  0.62  &  0.29  &  0.65  &  0.56  &  0.36  &  0.66  &  0.50  &  0.35  &  0.65  \\
Non-Energy Minerals &  2.00  &  0.47  &  2.01  &  1.99  &  1.82  &  2.00  &  2.04  &  1.88  &  1.96  &  1.97  & {\colorred\textbf{ 0.39 }} &  1.99  &  1.96  &  1.55  &  1.96  &  1.92  &  1.50  &  1.92  &  2.03  &  0.56  &  2.04  &  2.04  &  1.50  &  2.03  &  2.04  &  1.39  &  2.04  \\
Process Industries &  0.54  & {\colorred\textbf{ 0.31 }} &  0.65  &  0.58  &  0.51  &  0.65  &  0.55  &  0.46  &  0.65  &  0.58  &  0.31  &  0.67  &  0.56  &  0.47  &  0.67  &  0.58  &  0.47  &  0.67  &  0.55  &  0.33  &  0.65  &  0.56  &  0.47  &  0.66  &  0.53  &  0.42  &  0.66  \\
Producer Manufacturing &  0.77  & {\colorred\textbf{ 0.30 }} &  0.75  &  0.61  &  0.46  &  0.76  &  0.58  &  0.46  &  0.75  &  0.79  &  0.55  &  0.77  &  0.64  &  0.47  &  0.78  &  0.78  &  0.44  &  0.78  &  0.78  &  0.39  &  0.76  &  0.66  &  0.44  &  0.78  &  0.61  &  0.41  &  0.77  \\
Retail Trade &  0.46  & {\colorred\textbf{ 0.31 }} &  0.45  &  0.44  &  0.43  &  0.43  &  0.43  &  0.42  &  0.42  &  0.49  &  0.61  &  0.49  &  0.49  &  0.43  &  0.49  &  0.43  &  0.42  &  0.48  &  0.46  &  0.39  &  0.45  &  0.43  &  0.42  &  0.45  &  0.42  &  0.42  &  0.43  \\
Technology Services &  0.96  &  0.98  &  0.96  &  0.97  &  0.68  &  0.96  &  0.96  &  0.76  &  0.96  &  0.97  & {\colorred\textbf{ 0.34 }} &  0.97  &  0.98  &  0.67  &  0.96  &  0.98  &  0.66  &  0.98  &  0.98  &  0.69  &  0.97  &  0.99  &  0.68  &  0.97  &  0.97  &  0.66  &  0.97  \\
Transportation &  0.50  & {\colorred\textbf{ 0.39 }} &  0.50  &  0.49  &  0.48  &  0.49  &  0.57  &  0.52  &  0.50  &  0.51  &  0.41  &  0.52  &  0.51  &  0.48  &  0.51  &  0.51  &  0.47  &  0.52  &  0.97  &  0.40  &  0.50  &  0.51  &  0.45  &  0.49  &  0.49  &  0.44  &  0.50  \\
Utilities &  0.40  & {\colorred\textbf{ 0.28 }} &  0.39  &  0.36  &  0.36  &  0.37  &  0.39  &  0.36  &  0.39  &  0.41  &  0.30  &  0.41  &  0.37  &  0.36  &  0.40  &  0.40  &  0.36  &  0.41  &  0.40  &  0.28  &  0.39  &  0.37  &  0.34  &  0.38  &  0.39  &  0.33  &  0.40  \\
%%%%
%%%%
\toprule
\multirow{3}{*}{$\tau=0.1\%$} & \multicolumn{9}{c}{GED} & \multicolumn{9}{c}{Gaussian} & \multicolumn{9}{c}{Student--t}\\
\cmidrule(lr){2-10}\cmidrule(lr){11-19}\cmidrule(lr){20-28}
 & \multicolumn{3}{c}{EGARCH} & \multicolumn{3}{c}{GJR-GARCH} & \multicolumn{3}{c}{GARCH} & \multicolumn{3}{c}{EGARCH} & \multicolumn{3}{c}{GJR-GARCH} & \multicolumn{3}{c}{GARCH} & \multicolumn{3}{c}{EGARCH} & \multicolumn{3}{c}{GJR-GARCH} & \multicolumn{3}{c}{GARCH} \\
 \cmidrule(lr){2-4}\cmidrule(lr){5-7}\cmidrule(lr){8-10}\cmidrule(lr){11-13}\cmidrule(lr){14-16}\cmidrule(lr){17-19}\cmidrule(lr){20-22}\cmidrule(lr){23-25}\cmidrule(lr){26-28}
  & $\mathcal{M}_{\sS\sE}$ & $\mathcal{M}_{\sI\sV}$ & $\mathcal{M}_{\sN}$ & $\mathcal{M}_{\sS\sE}$ & $\mathcal{M}_{\sI\sV}$ & $\mathcal{M}_{\sN}$ & $\mathcal{M}_{\sS\sE}$ & $\mathcal{M}_{\sI\sV}$ & $\mathcal{M}_{\sN}$ & $\mathcal{M}_{\sS\sE}$ & $\mathcal{M}_{\sI\sV}$ & $\mathcal{M}_{\sN}$ & $\mathcal{M}_{\sS\sE}$ & $\mathcal{M}_{\sI\sV}$ & $\mathcal{M}_{\sN}$ & $\mathcal{M}_{\sS\sE}$ & $\mathcal{M}_{\sI\sV}$ & $\mathcal{M}_{\sN}$ & $\mathcal{M}_{\sS\sE}$ & $\mathcal{M}_{\sI\sV}$ & $\mathcal{M}_{\sN}$ & $\mathcal{M}_{\sS\sE}$ & $\mathcal{M}_{\sI\sV}$ & $\mathcal{M}_{\sN}$ & $\mathcal{M}_{\sS\sE}$ & $\mathcal{M}_{\sI\sV}$ & $\mathcal{M}_{\sN}$ \\
\hline
Commercial services &  0.56  &  0.40  &  0.60  &  0.51  &  0.59  &  0.55  &  0.53  &  0.56  &  0.69  &  0.43  &  0.33  &  0.66  &  0.62  &  0.46  &  0.60  &  0.47  &  0.45  &  0.65  &  0.46  &  0.36  &  0.53  &  0.45  & {\colorred\textbf{ 0.29 }} &  0.55  &  0.64  &  0.30  &  0.49  \\
Communications &  0.87  &  0.51  &  0.86  &  1.12  &  0.93  &  1.12  &  1.16  &  0.90  &  1.16  &  0.98  &  0.59  &  0.92  &  0.98  &  0.92  &  0.98  &  1.00  &  0.91  &  1.03  &  0.82  & {\colorred\textbf{ 0.45 }} &  0.80  &  0.88  &  0.66  &  0.84  &  0.87  &  0.68  &  0.84  \\
Consumer durables &  0.50  &  0.08  &  0.49  &  0.64  &  0.54  &  0.64  &  0.65  &  0.27  &  0.68  &  0.56  &  0.18  &  0.56  &  0.59  &  0.29  &  0.61  &  0.53  &  0.31  &  0.57  &  0.44  & {\colorred\textbf{ 0.03 }} &  0.41  &  0.46  &  0.07  &  0.41  &  0.41  &  0.09  &  0.40  \\
Consumer non-durables &  0.23  &  0.24  &  0.23  &  0.19  &  0.21  &  0.21  &  0.19  &  0.21  &  0.23  &  0.28  &  0.26  &  0.28  &  0.23  &  0.22  &  0.27  &  0.26  &  0.22  &  0.28  &  0.17  &  0.10  &  0.16  &  0.16  &  0.16  &  0.17  &  0.18  & {\colorred\textbf{ 0.07 }} &  0.20  \\
Consumer services &  0.73  &  0.25  &  0.72  &  0.74  &  0.75  &  0.73  &  0.89  &  0.69  &  0.75  &  0.83  & {\colorred\textbf{ 0.22 }} &  0.82  &  0.85  &  0.67  &  0.84  &  0.81  &  0.68  &  0.82  &  0.65  &  0.29  &  0.62  &  0.71  &  0.52  &  0.65  &  0.71  &  0.50  &  0.67  \\
Distribution services &  1.33  &  1.02  &  1.32  &  1.34  &  1.32  &  1.34  &  1.36  &  1.26  &  1.34  &  1.38  & {\colorred\textbf{ 0.96 }} &  1.38  &  1.39  &  1.29  &  1.39  &  1.38  &  1.35  &  1.38  &  1.28  &  1.06  &  1.28  &  1.33  &  1.12  &  1.30  &  1.35  &  1.07  &  1.32  \\
Electronic Technology &  1.01  &  0.99  &  0.99  &  1.01  &  0.57  &  1.01  &  1.01  &  0.62  &  0.95  &  1.11  & {\colorred\textbf{ 0.07 }} &  1.10  &  1.11  &  0.58  &  1.10  &  1.12  &  0.58  &  1.10  &  0.93  &  0.20  &  0.88  &  0.96  &  0.39  &  0.88  &  0.93  &  0.38  &  0.88  \\
Energy Minerals &  0.62  &  0.33  &  0.76  &  0.64  &  0.62  &  0.76  &  0.63  &  0.60  &  0.75  &  0.80  &  0.40  &  0.84  &  0.71  &  0.63  &  0.83  &  0.72  &  0.61  &  0.83  &  0.64  &  0.32  &  0.69  &  0.60  &  0.36  &  0.70  &  0.62  & {\colorred\textbf{ 0.31 }} &  0.70  \\
Finance &  0.47  &  0.20  &  0.47  &  0.53  &  0.52  &  0.51  &  0.54  &  0.51  &  0.49  &  0.55  &  0.53  &  0.54  &  0.58  &  0.52  &  0.57  &  0.60  &  0.51  &  0.55  &  0.38  & {\colorred\textbf{ 0.13 }} &  0.37  &  0.50  &  0.42  &  0.44  &  0.52  &  0.38  &  0.43  \\
Health Services &  0.34  &  0.15  &  0.64  &  0.48  &  0.44  &  0.67  &  0.42  &  0.43  &  0.67  &  0.39  &  0.20  &  0.70  &  0.43  &  0.48  &  0.71  &  0.54  &  0.48  &  0.74  &  0.33  & {\colorred\textbf{ 0.12 }} &  0.60  &  0.42  &  0.20  &  0.67  &  0.55  &  0.16  &  0.64  \\
Health Technology &  0.19  &  0.20  &  0.18  &  0.14  &  0.19  &  0.15  &  0.33  &  0.20  &  0.19  &  0.29  &  0.43  &  0.28  &  0.26  &  0.21  &  0.24  &  0.21  &  0.23  &  0.24  &  0.10  &  0.09  &  0.08  &  0.10  &  0.07  &  0.14  &  0.15  & {\colorred\textbf{ 0.03 }} &  0.14  \\
Industrial Services &  0.42  &  0.11  &  0.47  &  0.29  &  0.30  &  0.48  &  0.31  &  0.29  &  0.45  &  0.48  &  0.19  &  0.56  &  0.33  &  0.31  &  0.56  &  0.35  &  0.29  &  0.59  &  0.34  & {\colorred\textbf{ 0.09 }} &  0.38  &  0.25  &  0.14  &  0.40  &  0.23  &  0.11  &  0.39  \\
Non-Energy Minerals &  1.80  & {\colorred\textbf{ 0.16 }} &  1.82  &  1.79  &  1.63  &  1.80  &  2.02  &  1.71  &  1.72  &  1.86  &  0.24  &  1.88  &  1.85  &  1.30  &  1.85  &  1.79  &  1.24  &  1.79  &  1.74  &  0.16  &  1.74  &  1.76  &  0.72  &  1.74  &  1.76  &  0.50  &  1.75  \\
Process Industries &  0.37  &  0.16  &  0.54  &  0.42  &  0.43  &  0.55  &  0.38  &  0.36  &  0.54  &  0.49  &  0.23  &  0.61  &  0.46  &  0.39  &  0.61  &  0.50  &  0.38  &  0.62  &  0.28  & {\colorred\textbf{ 0.09 }} &  0.47  &  0.31  &  0.23  &  0.48  &  0.23  &  0.14  &  0.48  \\
Producer Manufacturing &  0.67  &  0.15  &  0.64  &  0.40  &  0.35  &  0.65  &  0.37  &  0.35  &  0.64  &  0.74  &  0.49  &  0.71  &  0.54  &  0.38  &  0.72  &  0.72  &  0.34  &  0.72  &  0.61  &  0.11  &  0.56  &  0.40  &  0.16  &  0.59  &  0.31  & {\colorred\textbf{ 0.10 }} &  0.59  \\
Retail Trade &  0.35  &  0.20  &  0.35  &  0.33  &  0.34  &  0.31  &  0.31  &  0.33  &  0.34  &  0.44  &  0.61  &  0.44  &  0.44  &  0.36  &  0.44  &  0.36  &  0.35  &  0.43  &  0.27  &  0.20  &  0.25  &  0.23  &  0.21  &  0.31  &  0.26  & {\colorred\textbf{ 0.20 }} &  0.29  \\
Technology Services &  0.85  &  0.85  &  0.84  &  0.86  &  0.58  &  0.84  &  0.85  &  0.69  &  0.84  &  0.92  & {\colorred\textbf{ 0.16 }} &  0.91  &  0.92  &  0.58  &  0.90  &  0.93  &  0.56  &  0.92  &  0.80  &  0.46  &  0.78  &  0.84  &  0.42  &  0.78  &  0.77  &  0.38  &  0.79  \\
Transportation &  0.39  &  0.25  &  0.39  &  0.38  &  0.39  &  0.38  &  0.56  &  0.45  &  0.38  &  0.46  &  0.31  &  0.46  &  0.46  &  0.41  &  0.46  &  0.45  &  0.40  &  0.46  &  0.79  &  0.19  &  0.31  &  0.35  &  0.21  &  0.31  &  0.31  & {\colorred\textbf{ 0.18 }} &  0.31  \\
Utilities &  0.31  &  0.19  &  0.30  &  0.27  &  0.28  &  0.27  &  0.30  &  0.29  &  0.30  &  0.37  &  0.25  &  0.36  &  0.32  &  0.30  &  0.36  &  0.35  &  0.30  &  0.36  &  0.25  &  0.15  &  0.24  &  0.21  &  0.14  &  0.25  &  0.26  & {\colorred\textbf{ 0.13 }} &  0.29  \\
\hline
\end{tabular}}
\caption{\footnotesize{Maximum AD in percentage points at confidence level $\tau=1\%$. For each sector, in {\colorred\textbf{bold}} we denote the minimum AD.}}
\label{tab:AD_all}
%\end{table}
\end{sidewaystable}

%::::::::::::::::::::::::::::::::::::::::::::::::::::::::::::::::::::::
% TABLE: KUPIEC 1%
%::::::::::::::::::::::::::::::::::::::::::::::::::::::::::::::::::::::
\begin{sidewaystable}[htbp]
\captionsetup{font={small}, labelfont=sc}
%\begin{table}[ht]
\centering
 \smallskip
  \centering
   \resizebox{1.0\columnwidth}{!}{%
    \setlength\tabcolsep{1mm}
\begin{tabular}{lcccccccccccccccccccccccccccc}
    \toprule
 \multirow{3}{*}{$\tau=1\%$}& \multicolumn{9}{c}{GED} & \multicolumn{9}{c}{Gaussian} & \multicolumn{9}{c}{Student--t}\\
\cmidrule(lr){2-10}\cmidrule(lr){11-19}\cmidrule(lr){20-28}
 & \multicolumn{3}{c}{EGARCH} & \multicolumn{3}{c}{GJR-GARCH} & \multicolumn{3}{c}{GARCH} & \multicolumn{3}{c}{EGARCH} & \multicolumn{3}{c}{GJR-GARCH} & \multicolumn{3}{c}{GARCH} & \multicolumn{3}{c}{EGARCH} & \multicolumn{3}{c}{GJR-GARCH} & \multicolumn{3}{c}{GARCH} \\
 \cmidrule(lr){2-4}\cmidrule(lr){5-7}\cmidrule(lr){8-10}\cmidrule(lr){11-13}\cmidrule(lr){14-16}\cmidrule(lr){17-19}\cmidrule(lr){20-22}\cmidrule(lr){23-25}\cmidrule(lr){26-28}
  & $\mathcal{M}_{\sS\sE}$ & $\mathcal{M}_{\sI\sV}$ & $\mathcal{M}_{\sN}$ & $\mathcal{M}_{\sS\sE}$ & $\mathcal{M}_{\sI\sV}$ & $\mathcal{M}_{\sN}$ & $\mathcal{M}_{\sS\sE}$ & $\mathcal{M}_{\sI\sV}$ & $\mathcal{M}_{\sN}$ & $\mathcal{M}_{\sS\sE}$ & $\mathcal{M}_{\sI\sV}$ & $\mathcal{M}_{\sN}$ & $\mathcal{M}_{\sS\sE}$ & $\mathcal{M}_{\sI\sV}$ & $\mathcal{M}_{\sN}$ & $\mathcal{M}_{\sS\sE}$ & $\mathcal{M}_{\sI\sV}$ & $\mathcal{M}_{\sN}$ & $\mathcal{M}_{\sS\sE}$ & $\mathcal{M}_{\sI\sV}$ & $\mathcal{M}_{\sN}$ & $\mathcal{M}_{\sS\sE}$ & $\mathcal{M}_{\sI\sV}$ & $\mathcal{M}_{\sN}$ & $\mathcal{M}_{\sS\sE}$ & $\mathcal{M}_{\sI\sV}$ & $\mathcal{M}_{\sN}$ \\
\hline
Commercial services & 0.38 & 4.36 & 0.21 & 1.12 & {\colorred\textbf{17.67}} & 0.38 & 0.38 & {\colorred\textbf{58.15}} & 0.21 & {\colorred\textbf{9.27}} & {\colorred\textbf{9.27}} & {\colorred\textbf{17.67}} & {\colorred\textbf{23.47}} & {\colorred\textbf{12.97}} & {\colorred\textbf{30.44}} & {\colorred\textbf{9.27}} & {\colorred\textbf{80.99}} & {\colorred\textbf{23.47}} & 1.82 & {\colorred\textbf{6.45}} & 1.82 & 2.86 & 0.21 & {\colorred\textbf{9.27}} & {\colorred\textbf{38.59}} & 0.21 & {\colorred\textbf{9.27}} \\
\bottomrule
\end{tabular}}
\caption{\footnotesize{p--values of the unconditional coverage test of Kupiec \citeyearpar{kupiec.1995} for the confidence level $\tau=1\%$. For each sector, in {\colorred\textbf{bold}} we denote the rejections of the null hypothesis at 5\% confidence level.}}
\label{tab:UC_test_kupiec}
%\end{table}
\end{sidewaystable}

%::::::::::::::::::::::::::::::::::::::::::::::::::::::::::::::::::::::
% TABLE: Christoffersen 1%
%::::::::::::::::::::::::::::::::::::::::::::::::::::::::::::::::::::::
\begin{sidewaystable}[!ht]
\captionsetup{font={small}, labelfont=sc}
%\begin{table}[ht]
 \smallskip
  \centering
   \resizebox{1.0\columnwidth}{!}{%
    \setlength\tabcolsep{1mm}
\begin{tabular}{lcccccccccccccccccccccccccccc}
    \toprule
\multirow{3}{*}{$\tau=1\%$} & \multicolumn{9}{c}{GED} & \multicolumn{9}{c}{Gaussian} & \multicolumn{9}{c}{Student--t}\\
\cmidrule(lr){2-10}\cmidrule(lr){11-19}\cmidrule(lr){20-28}
 & \multicolumn{3}{c}{EGARCH} & \multicolumn{3}{c}{GJR-GARCH} & \multicolumn{3}{c}{GARCH} & \multicolumn{3}{c}{EGARCH} & \multicolumn{3}{c}{GJR-GARCH} & \multicolumn{3}{c}{GARCH} & \multicolumn{3}{c}{EGARCH} & \multicolumn{3}{c}{GJR-GARCH} & \multicolumn{3}{c}{GARCH} \\
 \cmidrule(lr){2-4}\cmidrule(lr){5-7}\cmidrule(lr){8-10}\cmidrule(lr){11-13}\cmidrule(lr){14-16}\cmidrule(lr){17-19}\cmidrule(lr){20-22}\cmidrule(lr){23-25}\cmidrule(lr){26-28}
  & $\mathcal{M}_{\sS\sE}$ & $\mathcal{M}_{\sI\sV}$ & $\mathcal{M}_{\sN}$ & $\mathcal{M}_{\sS\sE}$ & $\mathcal{M}_{\sI\sV}$ & $\mathcal{M}_{\sN}$ & $\mathcal{M}_{\sS\sE}$ & $\mathcal{M}_{\sI\sV}$ & $\mathcal{M}_{\sN}$ & $\mathcal{M}_{\sS\sE}$ & $\mathcal{M}_{\sI\sV}$ & $\mathcal{M}_{\sN}$ & $\mathcal{M}_{\sS\sE}$ & $\mathcal{M}_{\sI\sV}$ & $\mathcal{M}_{\sN}$ & $\mathcal{M}_{\sS\sE}$ & $\mathcal{M}_{\sI\sV}$ & $\mathcal{M}_{\sN}$ & $\mathcal{M}_{\sS\sE}$ & $\mathcal{M}_{\sI\sV}$ & $\mathcal{M}_{\sN}$ & $\mathcal{M}_{\sS\sE}$ & $\mathcal{M}_{\sI\sV}$ & $\mathcal{M}_{\sN}$ & $\mathcal{M}_{\sS\sE}$ & $\mathcal{M}_{\sI\sV}$ & $\mathcal{M}_{\sN}$ \\
\hline
Commercial services &  1.31  & {\colorred\textbf{ 10.45 }} &  0.77  &  3.34  & {\colorred\textbf{ 22.82 }} &  1.31  &  1.31  & {\colorred\textbf{ 58.80 }} &  0.29  & {\colorred\textbf{ 18.91 }} & {\colorred\textbf{ 18.91 }} & {\colorred\textbf{ 22.82 }} & {\colorred\textbf{ 29.23 }} & {\colorred\textbf{ 17.27 }} & {\colorred\textbf{ 42.99 }} & {\colorred\textbf{ 12.65 }} & {\colorred\textbf{ 70.72 }} & {\colorred\textbf{ 36.57 }} & {\colorred\textbf{ 5.06 }} & {\colorred\textbf{ 14.28 }} & {\colorred\textbf{ 5.06 }} & {\colorred\textbf{ 7.40 }} &  0.77  & {\colorred\textbf{ 12.65 }} & {\colorred\textbf{ 9.28 }} &  0.29  & {\colorred\textbf{ 12.65 }} \\
\bottomrule
\end{tabular}}
\caption{\footnotesize{p--values of the conditional coverage test of Christoffersen \citeyearpar{christoffersen.1998} for $\tau=1\%$. For each sector, in {\colorred\textbf{bold}} we denote the rejections of the null hypothesis at 5\% confidence level.}}
\label{tab:CC_test_christoffersen}
%\end{table}
\end{sidewaystable}

%::::::::::::::::::::::::::::::::::::::::::::::::::::::::::::::::::::::
% TABLE: DQ 1%
%::::::::::::::::::::::::::::::::::::::::::::::::::::::::::::::::::::::
\begin{sidewaystable}[!ht]
\captionsetup{font={small}, labelfont=sc}
%\begin{table}[ht]
 \smallskip
  \centering
   \resizebox{1.0\columnwidth}{!}{%
    \setlength\tabcolsep{1mm}
\begin{tabular}{lcccccccccccccccccccccccccccc}
    \toprule
\multirow{3}{*}{$\tau=1\%$} & \multicolumn{9}{c}{GED} & \multicolumn{9}{c}{Gaussian} & \multicolumn{9}{c}{Student--t}\\
\cmidrule(lr){2-10}\cmidrule(lr){11-19}\cmidrule(lr){20-28}
 & \multicolumn{3}{c}{EGARCH} & \multicolumn{3}{c}{GJR-GARCH} & \multicolumn{3}{c}{GARCH} & \multicolumn{3}{c}{EGARCH} & \multicolumn{3}{c}{GJR-GARCH} & \multicolumn{3}{c}{GARCH} & \multicolumn{3}{c}{EGARCH} & \multicolumn{3}{c}{GJR-GARCH} & \multicolumn{3}{c}{GARCH} \\
 \cmidrule(lr){2-4}\cmidrule(lr){5-7}\cmidrule(lr){8-10}\cmidrule(lr){11-13}\cmidrule(lr){14-16}\cmidrule(lr){17-19}\cmidrule(lr){20-22}\cmidrule(lr){23-25}\cmidrule(lr){26-28}
  & $\mathcal{M}_{\sS\sE}$ & $\mathcal{M}_{\sI\sV}$ & $\mathcal{M}_{\sN}$ & $\mathcal{M}_{\sS\sE}$ & $\mathcal{M}_{\sI\sV}$ & $\mathcal{M}_{\sN}$ & $\mathcal{M}_{\sS\sE}$ & $\mathcal{M}_{\sI\sV}$ & $\mathcal{M}_{\sN}$ & $\mathcal{M}_{\sS\sE}$ & $\mathcal{M}_{\sI\sV}$ & $\mathcal{M}_{\sN}$ & $\mathcal{M}_{\sS\sE}$ & $\mathcal{M}_{\sI\sV}$ & $\mathcal{M}_{\sN}$ & $\mathcal{M}_{\sS\sE}$ & $\mathcal{M}_{\sI\sV}$ & $\mathcal{M}_{\sN}$ & $\mathcal{M}_{\sS\sE}$ & $\mathcal{M}_{\sI\sV}$ & $\mathcal{M}_{\sN}$ & $\mathcal{M}_{\sS\sE}$ & $\mathcal{M}_{\sI\sV}$ & $\mathcal{M}_{\sN}$ & $\mathcal{M}_{\sS\sE}$ & $\mathcal{M}_{\sI\sV}$ & $\mathcal{M}_{\sN}$ \\
\bottomrule
\end{tabular}}
%%%
\caption{\footnotesize{Out--of--sample Dynamic Quantile (DQ) test of Engle and Manganelli \citeyearpar{engle_manganelli.2004}, for $\tau=1\%$. For each sector, in {\colorred\textbf{bold}} we denote rejections of the null hypothesis at 5\% confidence level.}}
%%%
\label{tab:DQ_test_engle_manganelli}
%\end{table}
\end{sidewaystable}

%::::::::::::::::::::::::::::::::::::::::::::::::::::::::::::::::::::::
% TABLE: A/E MCS comparison
%::::::::::::::::::::::::::::::::::::::::::::::::::::::::::::::::::::::
%\begin{table}[!ht]
\begin{sidewaystable}[!ht]
\centering
   \resizebox{1.0\columnwidth}{!}{
\begin{tabular}{lccccccccccccc}
\hline
 & \multicolumn{6}{c}{A/E} & \multicolumn{6}{c}{AD} \\
  \cmidrule(lr){2-7}\cmidrule(lr){8-13}
 $\tau$& \multicolumn{3}{c}{$1\%$} & \multicolumn{3}{c}{$0.1\%$} & \multicolumn{3}{c}{$1\%$} & \multicolumn{3}{c}{$0.1\%$}\\
  \cmidrule(lr){1-1}\cmidrule(lr){2-7}\cmidrule(lr){8-13}
 Sector & $\mathcal{M}_{\mathsf{IV}}$ & $\mathcal{M}_{\mathsf{SE}}$ & $\mathcal{M}_{\mathsf{N}}$ & $\mathcal{M}_{\mathsf{IV}}$ & $\mathcal{M}_{\mathsf{SE}}$ & $\mathcal{M}_{\mathsf{N}}$ & $\mathcal{M}_{\mathsf{IV}}$ & $\mathcal{M}_{\mathsf{SE}}$ & $\mathcal{M}_{\mathsf{N}}$ & $\mathcal{M}_{\mathsf{IV}}$ & $\mathcal{M}_{\mathsf{SE}}$ & $\mathcal{M}_{\mathsf{N}}$\\
 \cmidrule(lr){1-1}\cmidrule(lr){2-4}\cmidrule(lr){5-7}\cmidrule(lr){8-10}\cmidrule(lr){11-13}
Commercial services &  0.72  &  --  &  --  &  1.63  &  --  &  --  &  0.53  &  --  &  --  &  0.34  &  --  &  -- \\
Communications &  0.75  &  --  &  --  &  1.00  &  --  &  --  &  1.01  &  --  &  --  &  0.68  &  --  &  -- \\
Consumer durables &  0.72  &  --  &  --  &  1.75  &  --  &  --  &  0.50  &  --  &  --  &  0.20  &  --  &  -- \\
Consumer non-durables &  0.78  &  0.79  &  0.89  &  3.00  &  2.33  &  2.50  &  0.28  &  0.31  &  0.32  &  0.18  &  0.19  &  0.20 \\
Consumer services &  0.79  &  0.83  &  0.87  &  2.71  &  2.60  &  2.96  &  0.80  &  0.89  &  0.89  &  0.64  &  0.71  &  0.69 \\
Distribution services &  0.83  &  --  &  --  &  2.58  &  --  &  --  &  1.37  &  --  &  --  &  1.23  &  --  &  -- \\
Electronic Technology &  0.71  &  0.88  &  0.90  &  2.75  &  2.60  &  2.35  &  0.69  &  1.17  &  1.16  &  0.52  &  0.97  &  0.92 \\
Energy Minerals &  0.88  &  --  &  --  &  1.75  &  2.47  &  1.71  &  0.71  &  --  &  --  &  0.50  &  0.67  &  0.73 \\
Finance &  --  &  0.59  &  0.76  &  2.67  &  2.92  &  2.71  &  --  &  0.60  &  0.61  &  0.48  &  0.49  &  0.45 \\
Health Services &  0.51  &  --  &  --  &  2.21  &  2.00  &  1.54  &  0.59  &  --  &  --  &  0.36  &  0.42  &  0.65 \\
Health Technology &  --  &  0.75  &  0.83  &  1.71  &  2.10  &  2.57  &  --  &  0.32  &  0.31  &  0.16  &  0.14  &  0.16 \\
Industrial Services &  0.66  &  --  &  --  &  1.79  &  2.07  &  1.68  &  0.40  &  --  &  --  &  0.24  &  0.31  &  0.45 \\
Non-Energy Minerals &  0.63  &  --  &  --  &  1.80  &  --  &  --  &  1.50  &  --  &  --  &  1.16  &  --  &  -- \\
Process Industries &  0.84  &  --  &  --  &  2.34  &  2.75  &  2.81  &  0.47  &  --  &  --  &  0.27  &  0.35  &  0.54 \\
Producer Manufacturing &  0.61  &  --  &  --  &  1.00  &  --  &  --  &  0.45  &  --  &  --  &  0.10  &  --  &  -- \\
Retail Trade &  0.98  &  --  &  --  &  2.64  &  3.38  &  3.58  &  0.42  &  --  &  --  &  0.28  &  0.29  &  0.31 \\
Technology Services &  0.69  &  0.82  &  0.84  &  3.17  &  2.63  &  2.46  &  0.68  &  0.97  &  0.97  &  0.54  &  0.81  &  0.81 \\
Transportation &  1.10  &  --  &  --  &  3.04  &  1.94  &  2.00  &  0.48  &  --  &  --  &  0.34  &  0.36  &  0.34 \\
Utilities &  1.33  &  --  &  --  &  2.84  &  3.15  &  3.17  &  0.35  &  --  &  --  &  0.22  &  0.26  &  0.27 \\
\bottomrule
\end{tabular}}
\caption{\footnotesize{A/E exceedances ratio and maximum AD averaged across the optimal MCS, $\hat{\rm M}_{1-\alpha}^*$, grouped by regressors and confidence levels: \qmo --\qmcsp means that no models  remained in the MCS at the end of the procedure. The maximum AD values are reported in percentage points.}}
\label{tab:MCS_average_AE_AD}
%\end{table}
\end{sidewaystable}

%::::::::::::::::::::::::::::::::::::::::::::::::::::::::::::::::::::::
% TABLE: AD MCS comparison
%::::::::::::::::::::::::::::::::::::::::::::::::::::::::::::::::::::::
\begin{sidewaystable}[!ht]
\centering
 \resizebox{1\columnwidth}{!}{
\begin{tabular}{lccccccccc}
\hline
 & \multicolumn{4}{c}{ADMax} & \multicolumn{4}{c}{ADMean}\\
 \cmidrule(lr){2-5}\cmidrule(lr){6-9}
 $\tau$& \multicolumn{2}{c}{$1\%$} & \multicolumn{2}{c}{$0.1\%$} & \multicolumn{2}{c}{$1\%$} & \multicolumn{2}{c}{$0.1\%$}\\
 \cmidrule(lr){1-1}\cmidrule(lr){2-3}\cmidrule(lr){4-5}\cmidrule(lr){6-7}\cmidrule(lr){8-9}
 Sector & $\mathrm{VaR_{dyn}}$ & $\mathrm{VaR_{avg}}$ & $\mathrm{VaR_{dyn}}$ & $\mathrm{VaR_{avg}}$ & $\mathrm{VaR_{dyn}}$ & $\mathrm{VaR_{avg}}$ & $\mathrm{VaR_{dyn}}$ & $\mathrm{VaR_{avg}}$\\
 %\cmidrule(lr){1-1}\cmidrule(lr){2-3}\cmidrule(lr){4-5}
 \hline
Commercial services &  --  &  --  &  0.511  &  0.877  &  --  &  --  &  0.133  &  0.247 \\
Communications &  1.010  &  1.664  &  0.995  &  1.516  &  0.183  &  0.367  &  0.217  &  0.250 \\
Consumer durables &  --  &  --  &  0.546  &  1.038  &  --  &  --  &  0.168  &  0.253 \\
Consumer non-durables &  0.299  &  0.774  &  0.300  &  0.577  &  0.085  &  0.362  &  0.085  &  0.244 \\
Consumer services &  0.851  &  1.372  &  0.844  &  1.203  &  0.145  &  0.366  &  0.147  &  0.249 \\
Distribution services &  1.333  &  1.887  &  1.378  &  1.733  &  0.119  &  0.364  &  0.124  &  0.246 \\
Electronic Technology &  0.956  &  1.722  &  0.934  &  1.553  &  0.197  &  0.369  &  0.209  &  0.252 \\
Energy Minerals &  0.728  &  1.356  &  0.745  &  1.217  &  0.148  &  0.370  &  0.156  &  0.255 \\
Finance &  0.600  &  1.145  &  0.608  &  1.006  &  0.142  &  0.364  &  0.164  &  0.248 \\
Health Services &  --  &  --  &  0.646  &  0.828  &  --  &  --  &  0.158  &  0.246 \\
Health Technology &  0.287  &  0.845  &  0.285  &  0.677  &  0.086  &  0.363  &  0.085  &  0.246 \\
Industrial Services &  --  &  --  &  0.453  &  0.818  &  --  &  --  &  0.146  &  0.250 \\
Non-Energy Minerals &  1.490  &  2.661  &  1.461  &  2.513  &  0.219  &  0.379  &  0.344  &  0.266 \\
Process Industries &  --  &  --  &  0.450  &  0.868  &  --  &  --  &  0.121  &  0.248 \\
Producer Manifacturing &  0.436  &  1.065  &  0.415  &  0.898  &  0.107  &  0.366  &  0.118  &  0.249 \\
Retail Trade &  0.450  &  0.977  &  0.428  &  0.810  &  0.110  &  0.366  &  0.106  &  0.249 \\
Technology Services &  0.741  &  1.472  &  --  &  --  &  0.201  &  0.366  &  --  &  -- \\
Transportation &  --  &  --  &  0.453  &  0.873  &  --  &  --  &  0.087  &  0.249 \\
Utilities &  0.346  &  0.880  &  0.338  &  0.712  &  0.067  &  0.365  &  0.082  &  0.248 \\
\bottomrule
\end{tabular}}
\caption{\footnotesize{Comparison between the dynamic VaR combination $\mathrm{VaR_{dynamic}}$ and the static average $\mathrm{VaR_{avg}}$ in terms of mean and maximum AD at both the considered confidence level $\tau=1\%$ and $\tau=0.1\%$. All values expressed in percentage points. The symbol \qmo --\qmcsp means that there remains only one model in the SSM.}}
\label{tab:var_comb_comparison}
\end{sidewaystable}

%
%::::::::::::::::::::::::::::::::::::::::::::::::::::::::::::::::::::::::::
% Section: Appendix C
%::::::::::::::::::::::::::::::::::::::::::::::::::::::::::::::::::::::::::
\newpage
\clearpage

%::::::::::::::::::::::::::::::::::::::::::::::::::::::::::::::::::::::
\section{Figures}
%::::::::::::::::::::::::::::::::::::::::::::::::::::::::::::::::::::::
%\label{sec:appendix_C}
%::::::::::::::::::::::::::::::::::::::::::::::::::::::::::::::::::::::
%

%::::::::::::::::::::::::::::::::::::::::::::::::::::::::::::::::::::::
% VaR 0.01 ALL MODELS
%::::::::::::::::::::::::::::::::::::::::::::::::::::::::::::::::::::::
%\begin{figure}[h]
%%
%\label{fig:VaR_ALL}
%%
%\centering
%  \begin{adjustbox}{addcode={\begin{minipage}{\width}}{\caption{%
%      \footnotesize{Time series plot of the log--returns \textit{(dark line)} along with the $\tau$--level VaR for $\tau=\left(1\%,0.1\%\right)$ for different sectors and different model specifications. In all panels the blue line denotes the ${\rm VaR}_{1\%}$ estimate for the EGARCH(1,1) model with no exogenous regressors, the red line denotes the ${\rm VaR}_{1\%}$ estimate for the EGARCH(1, 1) model with high, pos, neg, while the green and yellow lines denote the ${\rm VaR}_{0,1\%}$ for the model without and with exogenous informations, respectively.}}\end{minipage}},rotate=90,center}
%      %\begin{center}
%%
%\includegraphics[width=1.0\linewidth, height=0.66\linewidth]{Figures/VaR_TS_001_ALL_Assets_1fig.eps}
%%
%\end{adjustbox}
%%
%\end{figure}

%
%::::::::::::::::::::::::::::::::::::::::::::::::::::::::::::::::::::::
% VaR 0.01 ALL MODELS
%::::::::::::::::::::::::::::::::::::::::::::::::::::::::::::::::::::::

%%
%%::::::::::::::::::::::::::::::::::::::::::::::::::::::::::::::::::::::
%% VaR 0.01 MODELS SELECTED BY MSP
%% OF HANSEN 2011
%%::::::::::::::::::::::::::::::::::::::::::::::::::::::::::::::::::::::

%%
%%::::::::::::::::::::::::::::::::::::::::::::::::::::::::::::::::::::::
%% VaR 0.01 MODELS SELECTED BY MSP
%% OF HANSEN 2011
%%::::::::::::::::::::::::::::::::::::::::::::::::::::::::::::::::::::::
%\begin{figure}[ht]
%\centering
%  \begin{adjustbox}{addcode={\begin{minipage}{\width}}{\caption{%
%      \footnotesize{Time series plot of the log--returns \textit{(dark line)} along with the $\tau$--level VaR for $\tau=\left(1\%,0.1\%\right)$ for different sectors and different model specifications. In all panels the blue line denotes the ${\rm VaR}_{1\%}$ estimate for the EGARCH(1, 1) model with no exogenous regressors, the red line denotes the ${\rm VaR}_{1\%}$ estimate for the EGARCH(1, 1) model with high, pos, neg, while the green and yellow lines denote the ${\rm VaR}_{0,1\%}$ for the model without and with exogenous informations, respectively.}
%      }\end{minipage}},rotate=90,center}
%      %\begin{center}
%%
%\includegraphics[width=1.0\linewidth, height=0.66\linewidth]{Figures/VaR_TS_001_4Mod_Assets_1fig.eps}
%%
%\end{adjustbox}
%\label{fig:VaR_Dyn_Avg}
%\end{figure}

\clearpage



%% References

% SUBSECTION: BIBLIOGRAPHY
%==============================================================
\bibliographystyle{model1-num-names}
\bibliography{<your-bib-database>}






% \end{thebibliography}

%
\end{document}